\documentclass[11pt,a4paper,twoside,fleqn]{article}
\usepackage{color,graphicx,natbib,lscape,here,geometry,setspace,multirow,enumitem}
\usepackage[dvipsnames]{xcolor}
\usepackage{graphicx}
\usepackage{authblk}
\usepackage{silence}
\usepackage{helvet}

\usepackage{rotating}
\usepackage{multirow}

\usepackage{booktabs}
\usepackage{diagbox}
\usepackage{color}
\usepackage{setspace}
\usepackage{algorithm2e}
\usepackage{enumitem}
\usepackage{tikz}

\definecolor{nblue}{HTML}{000660}
\usepackage[colorlinks=true,urlcolor=nblue,linkcolor=nblue,citecolor=nblue]{hyperref}
\usepackage{bigstrut}

\usepackage{tabularx}
\usepackage{threeparttable}
\usepackage{dcolumn}
\newcolumntype{d}[1]{D{.}{.}{#1}}

\usepackage{array}
\newcolumntype{C}[1]{>{\centering\arraybackslash}p{#1}}

\usepackage{etoolbox} 
\makeatletter
\patchcmd{\BR@backref}{\newblock}{\newblock[}{}{}
\patchcmd{\BR@backref}{\par}{]\par}{}{}
\makeatother

\usepackage{pdflscape}
\usepackage{afterpage}

\usepackage{longtable}
\usepackage{array}

\usepackage{comment}
\usepackage{mathtools}

\usepackage[hang,flushmargin]{footmisc} 

\usepackage[hang]{footmisc}
\usepackage[british]{babel}

\pdfoutput=1
\usepackage{float}
\restylefloat{table}

\usepackage[title,titletoc]{appendix}
\makeatletter

\renewenvironment{appendices}{%
    \begin{oldappendices}%
    \renewcommand{\thefigure}{\ifnum \c@section>\z@ \thesection.\fi\@arabic\c@figure}%
    \@addtoreset{figure}{section}%
    \renewcommand{\thetable}{\ifnum \c@section>\z@ \thesection.\fi\@arabic\c@table}%
    \@addtoreset{table}{section}}{%
    \end{oldappendices}%
}\makeatother

\usepackage{titlesec} 
\titleformat{\section}[block]{\bfseries\Large}{\thesection. }{0em}{} 
\titleformat{\subsection}[block]{\large}{\thesubsection. }{0em}{\itshape} 
\titleformat{\subsubsection}[block]{}{\thesubsubsection. }{0em}{\itshape} 

\let\natbibcitet\citet
\renewcommand\citet{\bibpunct{(}{)}{,}{a}{,}{,}\natbibcitet}
\let\natbibcitep\citep
\renewcommand\citep{\bibpunct{(}{)}{;}{a}{,}{;}\natbibcitep}
\newcommand{\bi}{\begin{itemize}}
\newcommand{\ei}{\end{itemize}}
\newcommand{\be}{\begin{equation}}
\newcommand{\ee}{\end{equation}}
\defcitealias{ieo14}{IEO, 2014}

\long\def\symbolfootnote[#1]#2{\begingroup%
\def\thefootnote{\fnsymbol{footnote}}\footnote[#1]{#2}\endgroup}

\makeatletter
\def\ubar#1{\underline{\sbox\tw@{$#1$}\dp\tw@\z@\box\tw@}}
\def\obar#1{\overline{\sbox\tw@{$#1$}\dp\tw@\z@\box\tw@}}
\makeatother

\usepackage{bm}
\usepackage{caption}
\usepackage{subcaption}

\makeatletter\let\p@subfigure\thefigure\makeatother

\floatstyle{plaintop}
\restylefloat{table}

\usepackage{fancyhdr} 
\usepackage{cleveref}
\crefname{chapter}{Chapter}{Chapters}
\crefname{section}{Section}{Sections}
\crefname{subsection}{Section}{Sections}
\crefname{subsubsection}{Section}{Sections}
\crefname{figure}{Figure}{Figures}
\crefname{table}{Table}{Tables}
\crefname{equation}{Equation}{Equations}
\crefname{appendix}{Appendix}{Appendices}
\crefname{appendices}{Appendix}{Appendices}
\crefname{appsec}{Appendix}{Appendices}

\usepackage{catoptions}
\makeatletter

\def\Autoref#1{%
  \begingroup
  \edef\reserved@a{\cpttrimspaces{#1}}%
  \ifcsndefTF{r@#1}{%
    \xaftercsname{\expandafter\testreftype\@fourthoffive}
      {r@\reserved@a}.\\{#1}%
  }{%
    \ref{#1}%
  }%
  \endgroup
}
\def\testreftype#1.#2\\#3{%
  \ifcsndefTF{#1autorefname}{%
    \def\reserved@a##1##2\@nil{%
      \uppercase{\def\ref@name{##1}}%
      \csn@edef{#1autorefname}{\ref@name##2}%
      \autoref{#3}%
    }%
    \reserved@a#1\@nil
  }{%
    \autoref{#3}%
  }%
}
\makeatother

\title{\LARGE{\textbf{On the effectiveness of the European Central Bank's conventional and unconventional policies under uncertainty}}}
\author{\large{
Niko \uppercase{Hauzenberger}, Michael \uppercase{Pfarrhofer} and Anna \uppercase{Stelzer}}\thanks{
\noindent \textit{Corresponding author}: Anna Stelzer. Salzburg Centre of European Union Studies, University of Salzburg. \textit{Address}: M\"{o}nchsberg 2, 5020 Salzburg, Austria. \textit{Email}: \href{mailto:anna.stelzer@sbg.ac.at}{anna.stelzer@sbg.ac.at}. We thank Martin Feldkircher and Florian Huber for valuable comments and suggestions, and Guiseppe Ragusa for providing us with an update of the conventional and unconventional monetary policy factors extracted from the monetary policy event-study database (EA-MPD). All authors gratefully acknowledge financial support from the Austrian Science Fund (FWF, grant no. ZK $35$). Hauzenberger acknowledges funding by the Oesterreichische Nationalbank (project no. $18127$), Pfarrhofer also acknowledges funding by the Oesterreichische Nationalbank (project no. $18304$), and Stelzer acknowledges financial support by funds of the Humer Foundation.}
\\\vspace*{-0.5em}
\textit{University of Salzburg}}
\date{}


\def\equationautorefname~#1\null{%
  Eq.~(#1)\null
}
\def\equationautorefname~#1\null{
Eq.~(#1)\null
}

\usepackage[utf8, latin1]{inputenc}                 

\begin{document}
\setcounter{page}{1}\maketitle\thispagestyle{empty}\normalsize\vspace*{-2em}\small

\begin{center}
\begin{minipage}{0.8\textwidth}
\noindent\small In this paper, we investigate the effectiveness of conventional and unconventional monetary policy measures by the European Central Bank (ECB) conditional on the prevailing level of uncertainty. To obtain exogenous variation in central bank policy, we rely on high-frequency surprises in financial market data for the euro area (EA) around policy announcement dates. We trace the dynamic effects of shocks to the short-term policy rate, forward guidance and quantitative easing on several key macroeconomic and financial quantities alongside survey-based measures of expectations. For this purpose, we propose a Bayesian smooth-transition vector autoregression (ST-VAR). Our results suggest that transmission channels are impaired when uncertainty is elevated. While conventional monetary policy is less effective during such periods, and sometimes also forward guidance, quantitative easing measures seem to work comparatively well in uncertain times. 
\end{minipage}
\end{center}

\setstretch{1.1}\normalsize\renewcommand{\thepage}{\arabic{page}}
\newpage

\section{Introduction}\label{sec:introduction}
Conventional and unconventional monetary policies are widely used tools adopted in central banks to stimulate slacking economies and counteract disinflationary pressures stemming from recessionary episodes and economic uncertainty.\footnote{For a discussion of the relationship between uncertainty and conventional demand shocks that are typically counteracted by monetary policy, see \citet{leduc2016uncertainty}.} Research measuring the effects and transmission channels of monetary policy to the real and financial sectors has produced a voluminous body of literature \citep[see, among many others,][]{kashyap2000,kuttner2001monetary,bernanke2005measuring,gurkaynak2005actions,sims2006were,tenreyro2016pushing,altavilla2019measuring,jarocinski2020deconstructing}. In this paper, we contribute to this line of work by assessing non-linear features in monetary transmission for the euro area (EA). We consider economic uncertainty as a key determinant of the effectiveness of policy measures enacted by the European Central Bank (ECB), and discuss factors that may render policy interventions in uncertain times less effective.

A change in the monetary policy stance can be a direct reaction of the central bank to worsened economic conditions due to increases in uncertainty (e.g., lower or delayed investment and hirings when facing uncertainty shocks, see, for instance, \citealp{bloom2009}) via changes in the policy (target) rate, or more direct and long-term market interventions like quantitative easing (QE). It is worth mentioning that not only domestic uncertainty, but also uncertainty-spillovers from important globalized economies may play a role in this context \citep[see, e.g.,][]{davidson2019measuring}.

Monetary policy can also have more subtle effects when central banks communicate their decisions and actions to the public, a phenomenon the recent literature has also referred to as so-called information effects \citep{nakamura2018high,bauer2020fed,jarocinski2020deconstructing}.\footnote{By revealing otherwise concealed information (for instance, internal staff projections by the central bank regarding the economic outlook) to consumers and financial market participants, the central bank affects and causes updates to public expectations through policy announcements.} By transparently communicating the future path of policies and the decision makers' intentions, known as forward guidance (FG), central banks not only can reduce uncertainty about their future policy, they can also manage expectations of economic actors \citep{woodford2005}. Thereby, they ensure a more effective transmission of monetary policy. \citet{hutchinsonsmets2017} point out how crucial the ECB's clear communication of its policy reaction function to the public was in order to improve financial and economic conditions after a period of high uncertainty in the context of the European debt crisis. \citet{bekeartetal2013} also find that a looser monetary policy stance can reduce risk aversion and uncertainty. It is however not clear \textit{a priori} whether monetary policy works differently in times of high uncertainty when compared to low uncertainty environments.

Some authors have investigated asymmetric effects of monetary policy during expansions and recessions \citep[for an example, see][]{tenreyro2016pushing}. While high uncertainty levels and recessions often coincide, it is important to make a distinction between the two in the discussion of monetary policy transmission. Uncertainty can be elevated due to endogenous responses to the state of the business cycle \citep[see][]{ludvigson2020uncertainty}. In addition, exogenous uncertainty shocks (such as the 9/11 terror attacks or, abstracting from mandatory lockdown measures, the COVID-19 pandemic) may themselves cause drops in economic activity.\footnote{\citet{arellanoetal2019} and \citet{caldaraetal2016} find contractionary effects of uncertainty shocks to be particularly large in light of tight financial conditions.} However, it is worth reiterating that uncertainty (and sluggish macroeconomic and financial dynamics) can also occur without the realization of a recession. 

Different circumstances surrounding an economic recession might dampen the effectiveness of monetary policy, for several reasons. First, monetary policy transmission can be impaired because financial and credit markets are dysfunctional or distressed \citep[see][]{alessandrimumtaz2019}. Second, persistently low interest rates as a remnant of the policy reaction to previous economic recessions can lead to smaller effects of changes in interest rates on aggregate demand and output due to the presence of economic headwinds and inherent non-linearities \citep[see][]{boriohofmann2017}. Third, economic agents may be more hesitant to make investment or hiring decisions in times of increased uncertainty, hoping that more precise information to decide on longer-term actions might be available at a later point in time \citep{bloom2009}. This risk aversion implies that economic agents and financial markets may be less responsive to changes in economic conditions, such as interest rates. Such conditions, especially in the context of the first reason, may be closely related to the respective level of economic uncertainty regardless of whether the economy is in expansion or recession.

This paper contributes to the existing literature on monetary policy transmission under uncertainty as follows. We empirically illuminate the transmission of monetary policy in times of high and low uncertainty in the EA. There are some studies for the case of the US \citep[see][]{aastveitetal2017, caggianoetal2017}, however, to our knowledge there is no such analysis for the EA. We extend the preceding literature by including both CMP and UMP measures in our analysis (that is, we assess the effects of shocks to the policy rate, forward guidance and quantitative easing). These measures have gained increasing importance during and since the Great Recession. We achieve identification by extracting exogenous variation in policies via the high-frequency instruments developed in \citet{altavilla2019measuring}. From an econometric perspective, we propose a smooth-transmission vector autoregression \citep[ST-VAR, see also][]{granger1993modelling,auerbach2012measuring}, combined with a highly flexible hierarchical shrinkage prior to obtain precise empirical inference. The econometric framework is designed to trace monetary policy transmission based on the prevailing level of uncertainty in the economy. In order to investigate different channels of monetary policy transmission, we include a set of key financial and macroeconomic variables in our model, alongside survey-based measures of expectations.

Our results suggest that conventional transmission channels are often dysfunctional in times of uncertainty. We observe time-variation governed by the uncertainty indicator in direct transmission to key interest rates, spreads and a stock market index. Similarly, both the real economy and expectations show different responses in low versus high uncertainty periods. The prevailing level of uncertainty also affects the persistence of the shocks. These findings may be explained by the fact that information effects of monetary policy play a more important role during uncertain episodes. Moreover, elevated uncertainty affects the formation of expectations of economic actors, thereby reducing the effectiveness of expectation-related transmission channels.

The remainder of the paper is organized as follows. Section \ref{sec:econometrics} introduces the econometric framework used to evaluate the effectiveness of monetary policy in times of high and low uncertainty. Section \ref{sec:data} explains how we measure exogenous variation in monetary policy and provides details on the dataset. Our empirical findings are discussed and contrasted with the previous literature in Section \ref{sec:results}. Section \ref{sec:conclusions} offers closing remarks. The Appendix contains further details on our estimation algorithm and the data.

\section{Econometric framework}\label{sec:econometrics}
\subsection{The smooth-transition vector autoregression}
Let $\bm{y}_t$ denote an $M\times1$-vector containing the series of interest at time $t=1,\hdots,T$. We assume a ST-VAR model of the form:
\begin{align}
    \bm{y}_t = &(\bm{A}_{11} \bm{y}_{t-1} + \hdots + \bm{A}_{1P} \bm{y}_{t-P} + \bm{c}_1 + \bm{\delta}_{1s} x_{st}) \times S_t(u_{t-1}) + \label{eq:benchmarkVAR}\\ & (\bm{A}_{01} \bm{y}_{t-1} + \hdots + \bm{A}_{0P} \bm{y}_{t-P} + \bm{c}_0 +\bm{\delta}_{0s} x_{st}) \times \left(1-S_t(u_{t-1})\right) + \bm{\epsilon}_t.\nonumber
\end{align}
where $\bm{A}_{ip}$ are $M\times M$-coefficient matrices for state $i\in\{0,1\}$ and lag $p=1,\hdots,P$, and $\bm{c}_i$ is an $M\times1$-vector of intercepts. The state indicators $S_t(u_{t-1})\in[0,1]$ transition smoothly between regimes and depend on a signal variable $u_{t-1}$ (in our case, a measure of uncertainty). They are bounded between zero and one and discussed in more detail below. $\bm{\epsilon}_t \sim \mathcal{N}(\bm{0},\bm{\Omega})$ is a Gaussian error term with zero mean and $M\times M$-covariance matrix $\bm{\Omega}$.\footnote{Earlier research on the EA has shown that evidence for time-variation in the volatility of key series is muted \citep[see, for instance,][]{jarocinski2018inflation}. Such features are mainly required for data from the United States, when jointly modeling distinct periods such as the Great Inflation (approximately 1965--1982) versus the Great Moderation (around 1985--2007), see, e.g., \citet{clark2011real}. Given the comparatively brief existence of the EA since 1999, this is not the case in the context of our application. Accordingly, we follow the recent literature on EA monetary policy analysis using a homoscedastic specification, see for instance, \citet{burriel2018uncovering} or \citet{jarocinski2020deconstructing}. For these reasons, we refrain from introducing regime-switching covariance matrices. Moreover, it is worth noting that our identification approach is independent of the covariance matrix (different to \citealp{auerbach2012measuring} or \citealp{caggiano2014uncertainty}), which implies that even though we rule out time-varying volatilities, the impacts of the shocks feature time-variation.}

We include CMP and UMP shocks (indexed $s\in\{\text{TG},\text{FG},\text{QE}\}$, with target, TG -- policy rate; forward guidance, FG; and quantiative easing, QE) one at a time as scalar exogenous instruments $x_{st}$ (for details how these instruments are constructed, see Subsection \ref{sec:instruments}). The $M\times1$-vector $\bm{\delta}_{is}$ measures the contemporaneous responses of the endogenous variables to the shock $s$, and is thus the state-specific impact vector that can be used for structural inference \citep[for details, see][]{paul2020time}.

The ST-VAR has several attractive properties when contrasted with related econometric approaches.
Compared to deterministic regime-classification, threshold or Markov switching (MS) models, with potentially only a small number of observations in one of the regimes, the ST-VAR informs its parameter estimates based on variation in the degree of particular regimes (loosely related to probabilities). This sharpens inference, and biases the coefficients towards a linear specification. When compared to conventional time-varying parameter (TVP) models with gradually evolving coefficients (that is, with a random walk state equation), using our specification with a pre-defined defined signal variable $u_{t-1}$ allows to link time-variation to observed factors.\footnote{Recent papers overcome this shortcoming of conventional TVP models by augmenting the state equations of the TVPs by latent or observed factors \citep{chan2020reducing,fischer2020flexible}. This allows for addressing why specific parameters change over time, similar to our interpretation of the signal variable driving changes in the parameters, and thus, impulse responses.}

A crucial modeling decision for the ST-VAR is both the choice of the signal variable $u_t$ and the transition function for $S_t(u_{t-1})$. We follow the related literature \citep[see, e.g.,][]{auerbach2012measuring,caggiano2014uncertainty} and choose:
\begin{equation*}
    S_t(u_{t-1}) = \frac{1}{1+\exp\left(-\phi(u_{t-1} - \gamma)\right)},
\end{equation*}
a first-order logistic function as the transition function, with $u_{t-1}$ denoting the first lag (to render the parameter $\phi$ scale-invariant, we standardize $u_t$ to have zero mean and unit variance prior to estimation, see also \citealp{gefang2009nonlinear}). We include the signal variable, a measure of uncertainty in line with our research focus, as first lag to avoid contemporaneous feedback from policy actions depending on whether we are in a low/high uncertainty regime. 

The parameter $\phi>0$ governs the speed of adjustment, while $\gamma$ marks a threshold value that separates the two regimes. Naturally, depending on the actual evolution of the signal variable $u_t$, the parameter $\gamma$ determines the split at which observations are allocated more to either state zero and one. The parameter $\phi$ governs how smoothly the economy transitions between states. In the limiting case with $\phi\to\infty$, the state indicator $S_t(u_{t-1})$ switches between zero and one (marking clearly separated regimes), closely related to conventional threshold VARs \citep[see, e.g.,][]{alessandrimumtaz2019,huber2019threshold}. For the case of $\phi\to0$, the logistic function turns constant, with $S_t(u_{t-1})=0.5$ effectively resulting in a linear VAR specification.

\subsection{Prior setup}\label{sec:priors}
\subsubsection{Priors for the state-specific VAR coefficients}
Our proposed prior setup for the autoregressive coefficients is similar to \citet{hauzenberger2020stochastic} who use a MS vector error correction model. While MS models produce a strictly binary regime allocation --- $S_t(u_{t-1})$ would either be zero or one at time $t$ --- our setup implies that the coefficient matrices are weighted averages across states. We design our prior to be centered on the corresponding linear specification of our model, and pool coefficients across regimes.

To achieve this, we collect the regime specific coefficients in $\bm{A}_i = (\bm{A}_{i1},\hdots,\bm{A}_{iP},\bm{c}_i,\bm{\delta}_i)$ for states $i\in\{0,1\}$, and construct $\bm{a}_i = \text{vec}(\bm{A}_i)$ of size $J=(M(Mp+2))\times1$. The prior variances are collected in $\tilde{\bm{V}}_i=\text{diag}\left(\{\tilde{v}_{ij}\}_{j=1}^{J}\right)$ and $\bm{V}=\text{diag}\left(\{{v}_{j}\}_{j=1}^{J}\right)$ with $\tilde{v}_{ij} = \tilde{\lambda}_i^2\tilde{\psi}_{ij}^2$ and ${v}_{j} = {\lambda^2}{\psi}_{j}^2$. We propose a hierarchical global-local shrinkage setup based on the horseshoe \citep[HS,][]{carvalho2010horseshoe} prior:\footnote{We choose the horseshoe prior due to its excellent shrinkage properties and the lack of a prior tuning hyperparameter. Note that any prior from the class of global-local shrinkage priors \citep[see, e.g.,][]{polson2010shrink} can be used.}
\begin{align}
    \bm{a}_i\sim\mathcal{N}(\tilde{\bm{a}},\tilde{\bm{V}}_i), \quad
    &\tilde{\lambda}_{i}\sim\mathcal{C}^{+}(0,1), \quad \tilde{\psi}_{ij}\sim\mathcal{C}^{+}(0,1),\label{eq:regimeVAR}\\
    \tilde{\bm{a}}\sim\mathcal{N}(\bm{a},\bm{V}), \quad
    &{\lambda}\sim\mathcal{C}^{+}(0,1), \quad {\psi}_{j}\sim\mathcal{C}^{+}(0,1),\label{eq:poolVAR}
\end{align}
for $i\in\{0,1\}$ and $j=1,\hdots,J$. The symbol $\mathcal{C}^{+}$ denotes the half-Cauchy distribution. Equation (\ref{eq:regimeVAR}) states that the regime-specific coefficients arise from a Gaussian distribution with mean $\tilde{\bm{a}}$, and shrinkage is governed by two regime-specific shrinkage parameters collected in $\tilde{\bm{V}}_i$.

The first is a global shrinkage parameter, $\tilde{\lambda}_{i}$, which pushes all coefficients in $\bm{a}_i$ towards $\tilde{\bm{a}}$, and thus, a linear model specification (the case of exact parameter homogeneity across regimes). The second type are local scaling parameters, $\tilde{\psi}_{ij}$, that allow for variable-specific deviations from linearity in each regime. Rather than choosing specific values for $\tilde{\bm{a}}$, we assign a Gaussian prior centered on $\bm{a}$ and estimate these coefficients, see Eq. (\ref{eq:poolVAR}). We choose a prior mean of $0.95$ for all first-order autoregressive coefficients if the respective series is transformed as log-levels, and set it to zero in all other cases (mimicking a Minnesota-type prior specification). Choosing $0.95$ rather than unity implies that our prior is centered on a strictly stationary multivariate model. Here, we introduce another hierarchy of shrinkage to regularize the high-dimensional parameter space. Again, we rely on the HS prior and introduce a global shrinkage parameter $\lambda$ alongside local scalings $\psi_j$.

Intuitively, our setup implies that we impose conventional shrinkage towards a stylized prior model on the linear specification's parameters in $\tilde{\bm{a}}$. In a second step, we impose regime-specific shrinkage of the non-linear coefficients towards the linear model.\footnote{In this regard, our prior is similar to the hierarchical cross-sectional pooling prior proposed for time-varying parameter global VARs in \citet{pfarrhofer2019measuring}.} Our setup differs from \citet{hauzenberger2020stochastic} based on our choice of the respective shrinkage prior. Introducing additional global shrinkage parameters by regime implies that our model is capable of detecting a regime-specific degree of shrinkage, which is particularly useful if observations are unevenly distributed across states.

\subsubsection{Priors for the covariance matrix}
Our model has a potentially huge-dimensional parameter space, which quickly becomes computationally prohibitive to sample in one block if the number of endogenous variables ($M$) or the lag order ($P$) are increased. To alleviate the computational burden, we rely on triangularizing the model following \citet{carriero2019large} to allow for equation-by-equation estimation. Appendix \ref{app:A} shows how to write the multivariate system of equations as a set of $M$ independent regressions. 

To establish our prior, we decompose the covariance matrix of the error $\bm{\epsilon}_t$ in Eq. (\ref{eq:benchmarkVAR}) as $\bm{\Omega} = \bm{H}\bm{\Sigma}\bm{H}'$ with $\bm{H}$ denoting the normalized lower Choleski factor of $\bm{\Omega}$ and $\bm{\Sigma} = \text{diag}\left(\{\sigma_m^2\}_{m=1}^M\right)$ is an $M\times M$ matrix with equation-specific variances $\sigma_m^2$ for $m=1,\hdots,M$ on its diagonal. 

We collect the free elements (those below the diagonal) of the matrix $\bm{H}$ in an $R=(M(M-1)/2)\times1$-vector $\bm{h}$ and index its elements by $h_r$ with $r=1,\hdots,R$. Here, we define a $R \times R$-matrix $\hat{\bm V} = \text{diag}\left(\{\hat{v}_{r}\}_{r=1}^{R}\right)$ with $\hat{v}_{r} = \hat{\lambda}^2\hat{\psi}_r^2$ and again impose a HS prior:
\begin{equation}
    \bm h \sim\mathcal{N}(0, \hat{\bm V}),\quad \hat{\lambda}\sim\mathcal{C}^{+}(0,1), \quad \hat{\psi}_{r}\sim\mathcal{C}^{+}(0,1).\label{eq:covVAR}
\end{equation}
Again, the global parameter $\hat{\lambda}$ imposes shrinkage towards zero for all free elements of the matrix $\bm{H}$, while the local parameters $\hat{\psi}_r$ allow for deviations of the $r$th coefficient.

On the diagonal elements of $\bm{\Sigma}$, we impose a set of $M$ weakly informative independent inverse Gamma priors:
\begin{equation*}
    \sigma^2_m\sim\mathcal{G}^{-1}(3,0.3),\quad \text{for } m=1,\hdots,M.
\end{equation*}

\subsubsection{Priors for the state transition function}
In our empirical application, we standardize the signal variable $u_t$ such that it has mean zero and unit variance. We construct the prior for the threshold parameter $\gamma$ such that it is bounded between the minimum and maximum values of $u_t$ and center it on zero which implies a prior centered on the mean of $u_t$ in the scale of the untransformed series:
\begin{equation*}
    \gamma\sim\mathcal{TN}\left(0,\sigma^2_{\gamma},\min(u_t),\max(u_t)\right)
\end{equation*}
We choose $\sigma^2_{\gamma}=0.01$ implying an informative prior. This prior information is needed to impose a sensible regime-allocation in our model. It is worth mentioning that while the prior pushes the parameter towards zero, likelihood information may still overrule the tightness of the prior, different to using hard-coded restrictions as in \citet{auerbach2012measuring}.

The speed of adjustment parameter follows an inverse Gamma distribution:
\begin{equation*}
    \phi\sim\mathcal{G}^{-1}(a^2/b,a/b).
\end{equation*}
The hyperparameters are chosen such that $a$ denotes the mean of the prior and $b$ its variance. We set $a=2$ with variance $b=0.01$. This implies that the economy spends about $25$ percent of the time in the high-uncertainty regime. A similar reasoning for introducing substantial prior information as for $\gamma$ applies here. Calibrating the prior in this manner imposes a sensible regime-allocation in our model, yet we still obtain a posterior distribution for the parameter and deviations from the prior are possible if likelihood information prevails in the respective posterior.

This completes our prior setup. Combining the priors with the likelihood of the model results in a set of mostly standard conditional posterior distributions that can be used for Gibbs sampling. For the parameters of the state transition function, we rely on a Metropolis-within-Gibbs step. These can be used for designing a Markov chain Monte Carlo (MCMC) algorithm to obtain posterior inference. Details on the resulting posterior distributions and the sampling algorithm are provided in Appendix \ref{app:B}.

\section{Data and model specification}\label{sec:data}
\subsection{Measuring monetary policy shocks and identification}\label{sec:instruments}
We state in Eq. (\ref{eq:benchmarkVAR}) that monetary policy shocks are included in our model as exogenous instruments $x_{st}$. In this paper, we assess both CMP and UMP shocks following the methodology set forth in \citet{altavilla2019measuring}, relying on the \textit{Euro Area Monetary Policy Event-Study Database} (EA-MPD).\footnote{Raw event-study data is available at \href{https://www.ecb.europa.eu/pub/pdf/annex/Dataset_EA-MPD.xlsx}{ecb.europa.eu/pub/pdf/annex/Dataset\_EA-MPD.xlsx}. The dataset is described in detail in \citet{altavilla2019measuring}.} Before discussing how the instruments are established, we provide a brief overview on ECB communication that can be exploited for high-frequency identification.\footnote{An early paper on high-frequency identification of monetary policy shocks using tick-frequency asset price data is \citet{kuttner2001monetary}. Related methods have subsequently been used for monetary policy analysis, with some examples covering both the US and the EA given by \citet{bernanke2005explains,gurkaynak2005actions,brand2010impact,andrade2020delphic,hauzenberger2019bayesian} and \citet{jarocinski2020deconstructing}.}

ECB policy decisions are typically communicated to the public in a two-step process. First, there is a press release which provides information on the policy decision (without discussing how the Governing Council came to this decision). Second, the press release is followed by a press conference 45 minutes later, where the President reads a prepared statement explaining the previously announced policies (which takes about 15 minutes), and answers questions in a Q\&A session afterwards (about 45 minutes). The latter is often informative about the ECB's outlook and the future path of monetary policy. In conjunction, these two ``monetary events'' and precise time-stamps are employed to identify different policy surprises on financial markets and thus provide a measure of exogenous variation in monetary policy. 

The EA-MPD collects high-frequency market reactions in narrow windows (ten minutes before and after the event) around both monetary events, based on tick data for interest rates at different maturities and stock returns, among others. In essence, if monetary policy actions by the ECB were fully anticipated by market participants, we should observe no reaction of asset prices during either the press release or the press conference. If, however, the ECB communicates unanticipated actions (that are thus orthogonal to the information set of the public), the market would quickly adjust and price this new information. \citet{altavilla2019measuring} exploit these dynamics in a wide range of asset classes using an identified factor model (subject to orthogonality restrictions), and pin down four different shocks: the \textit{target}, \textit{timing}, \textit{forward guidance} and \textit{quantitative easing} shock.\footnote{For details on identification, see also \citet{gurkaynak2005actions}, \citet{brand2010impact} and \citet{swanson2020measuring}.}

In this paper, we rely on these factors to trace the transmission of both CMP and UMP shocks to the real and financial economy in times of uncertainty. We select the target (TG), forward guidance (FG) and quantitative easing (QE) as our shocks of interest and include them in our model as exogenous instruments:\footnote{The timing shock is very similar to the forward guidance shock, with the only difference being the time horizon for which forward guidance is applicable. To economize on space, we focus on the forward guidance instrument aimed at the two-year ahead horizon discussed below.}
\begin{itemize}[leftmargin=1cm,labelsep=0.65cm]
    \item \textit{Target} (TG): The TG shock is derived from announcements during the press release. This factor does not load on surprises during the press conference window, implying that the series captures mainly CMP. In terms of relevant instruments, the factor exhibits loadings on short-term yields, with a maximum loading on the one-month overnight index swap (OIS) rate.
    \item \textit{Forward guidance} (FG): The FG factor loads on series during the press conference window, and mostly affects the middle segment of the yield curve (with a peak effect at about two years of maturity and substantial loadings up to five years). The FG factor captures revisions in market expectations about the future path of monetary policy that are orthogonal to the TG factor's current policy surprise content. We interpret this shock series as the first measure of UMP.
    \item \textit{Quantitative easing} (QE): The QE shock is designed to dominate in the press conference window. This factor shows maximum loadings for yields with ten-year maturity, reflecting the long end of the yield curve. It is worth mentioning that based on the identifying restrictions of the factor model, QE is only present starting in $2014$, consistent with the historical evolution of UMP by the ECB. We interpret this shock as our second measure of UMP.
\end{itemize}

The exogenous shocks $x_{st}$ for $s\in\{\text{TG},\text{FG},\text{QE}\}$ are included one at a time in our model set forth in Eq. (\ref{eq:benchmarkVAR}), and the vectors $\bm{\delta}_{i,\text{TG}}$, $\bm{\delta}_{i,\text{FG}}$ and $\bm{\delta}_{i,\text{QE}}$ measure the sensitivity of the endogenous variables in $\bm{y}_t$ to these shocks in regime $i\in\{0,1\}$. Using the identified contemporaneous responses, we can calculate higher-order impulse response functions by tracing the impacts through the dynamic multivariate system. This approach to identifying the dynamic impact of monetary policy shocks is similar to \citet{gertler2015monetary} and \citet{paul2020time}.

An important aspect of monetary policy analysis using high-frequency data are the aforementioned central bank information effects \citep[see, for instance,][]{nakamura2018high,jarocinski2020deconstructing}.\footnote{Note that the derivation of the factors we use as instruments is unaffected by the presence of information shocks \citep{altavilla2019measuring}.} In a nutshell, it is argued that central bank announcements not only convey the respective policy decision, but also reveal information about the projected economic outlook. \citet{jarocinski2020deconstructing} investigate information effects in detail for the EA by assessing the high-frequency responses of three-month OIS rates in conjunction with Euro Stoxx 50 movements around announcement dates. They find that positive comovement in both surprises is indicative of an information shock, and impulse responses look substantially different when compared to those of a \textit{pure} monetary policy shock (with negative comovement of the surprise series, in line with theory). 

\citet{miranda2020transmission} propose a procedure for pre-processing instruments, by purging them from predictable components, to achieve clean identification of monetary policy shocks which yields similar results. By contrast, \citet{bauer2020fed} find that what \citet{nakamura2018high} call information shocks (capturing differences in public and private information on the economic outlook of the central bank) might actually be artefacts of economic agents and the central bank reacting to the same news shock. Given the close relationship between uncertainty and news shocks \citep{berger2020uncertainty}, this might be one of the channels why the effectiveness of monetary policy is affected during periods of elevated uncertainty.

While our econometric framework in principle allows for disentangling monetary policy from information shocks and tracing the impacts of both individually,\footnote{One could, for instance, employ the approach referred to as \textit{poor man's sign restrictions} by \citet{jarocinski2020deconstructing}. This implies that for months where high-frequency surprises move in the same direction, the factors would be set to zero. A major drawback of this strategy is that months are binarily classified as either \textit{pure} monetary policy or information shocks, ruling out cases in between.} we refrain from doing so, for three reasons: 
\begin{enumerate}[leftmargin=1cm,labelsep=0.65cm]
    \item We closely follow \citet{altavilla2019measuring}, who use the external instruments directly for identifying TG, FG and QE shocks in their financial VAR analysis. Our two-regime model is equipped for endogenously selecting distinct regimes, similar to their sub-sample analysis where they detect the presence of potential information shocks.
    \item The informational content and role of information shocks in the context of FG and QE is less clear than when focusing solely on TG shocks \citep[see the discussion in][]{bauer2020fed}. Putting specific restrictions on reactions may in fact bias estimates of the actual shocks.
    \item Our paper intends to shed light on the transmission of CMP and UMP in times of uncertainty. It seems a worthwhile empirical exercise to study the transmission of the actual measures of our exogenous shocks to the real and financial economy over time. And this transmission also includes the presence of information effects during policy announcements by the ECB. The information effects potentially impaired the effectiveness of policies during specific times, for instance, when uncertainty levels were elevated. We elaborate on this notion in the context of the discussion of our results below.
\end{enumerate}
In fact, \citet{bauer2020fed} argue that ``even though high-frequency monetary policy surprises may be correlated with macroeconomic data \textit{ex post}, they still can by used, without adjustment, to estimate the effects of an exogenous change in monetary policy [...].'' To ensure that our factors are valid external instruments for our shocks of interest, we follow \citet{miranda2020transmission} and project out \textit{ex post} correlation with the information set contained in the ST-VAR. We achieve this by regressing the instruments on its own $P$ lags and the contemporaneous values of the variables in $\bm{y}_t$ and $P$ of their lags, and use the residuals from this regression as the respective instrument.

\begin{figure}[t]
\includegraphics[width=\textwidth]{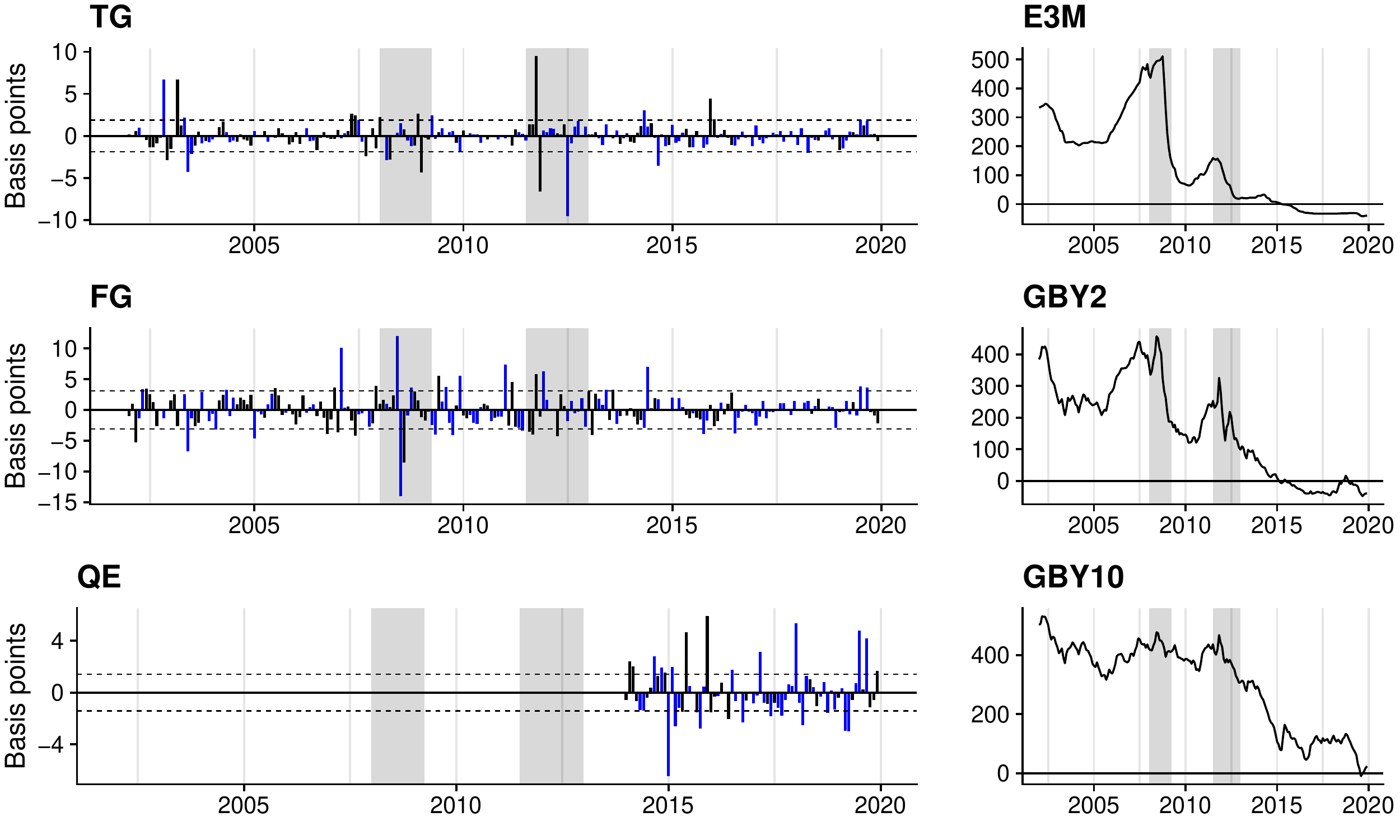}
\caption{Exogenous instruments (target, TG/forward guidance, FG/quantitative easing, QE), signaling effects and interest rates at different maturities.}\label{fig:factors}\vspace*{-0.3cm}
\caption*{\footnotesize\textit{Notes}: The bars indicate the respective value of the factor per month. Bars in blue show months where at least one policy announcement resulted in positive comovements of high-frequency surprises in three-month OIS rates and the Euro Stoxx 50 stock index (information shock). Black bars show negative comovements of these two high-frequency series (conventional monetary policy shock in line with theory), see also \citet{jarocinski2020deconstructing}. The dashed lines mark one standard deviation of the shocks. The grey shaded areas indicate recessions dated by the CEPR Euro Area Business Cycle Dating Committee.}
\end{figure}

Figure \ref{fig:factors} plots the exogenous instruments (TG, FG and QE) and corresponding interest rates (three-month Euribor, E3M; EA two-year government bond yields, GBY2; EA ten-year government bond yields, GBY10; see also next subsection) on a monthly frequency. The figure is adjusted to indicate months where information effects may have played a role during ECB policy announcements. Note that while this is especially important in the context of the TG factor, the role of information shocks is less clear regarding FG and QE shocks. The months with potential information shocks are marked in blue (based on at least one policy announcement during the respective month resulting in positive comovements of the three-month OIS rate and Euro Stoxx 50 surprises). 

Several findings are worth noting. First, the largest movements in both the TG and FG factor are visible during the two crisis episodes (the Great Recession and the subsequent European debt crisis). Second, while we observe positive comovement shocks throughout our sample, they occur most frequently during crisis episodes and after 2015. These crisis episodes are also often accompanied by elevated levels of uncertainty. While we discuss this aspect in more detail below in the context of our structural results, we conjecture that this may be one of the reasons why monetary policy during uncertain times appears to be less effective. Positive comovement shocks tend to cancel the effects of conventional monetary policy \citep[see][]{jarocinski2020deconstructing}. Third, the QE factor is only active after 2014. Here, we observe information shocks particularly around 2015 and at the end of the sample. Fourth, one striking observation when considering the corresponding interest rates is that rate cuts do not necessarily result in surprises in the instrument of the same sign. Positive surprises, for example, occur if financial markets expected larger cuts than realized. While the ECB intended to conduct expansionary policy, the less-than-expected cuts act like contractionary shocks.

\subsection{Real, financial and survey-based measures}
The vector of endogenous variables, $\bm{y}_t$, contains several macroeconomic and financial time series, alongside survey-based measures of expectations. In particular, we include the Euribor three-month rate (\texttt{E3M}) as our short-term target rate. To trace the effects of FG and QE shocks, we also consider two-year (\texttt{GBY2}, targeted by the FG factor) and ten-year (\texttt{GBY10}, targeted by the QE factor) EA government bond yields. Our set of financial indicators is completed by the Euro Stoxx 50 stock market index (\texttt{ES50}), and the ICE BofA Euro high-yield index option-adjusted spread (\texttt{OAS}) to measure financial conditions.

As key macroeconomic indicators, we include the harmonized index of consumer prices (\texttt{HICP}), the unemployment rate (\texttt{UNEMP}) and industrial production excluding construction (\texttt{IP}, as monthly indicator of economic activity). To gauge the impact of monetary policy shocks on firms and household expectations, we refer to the following survey-based measures. We include industrial (\texttt{ISICI}) and consumer confidence (\texttt{CSCCI}) indicators, consumers' unemployment expectations over the next twelve months (\texttt{CSU12}), and consumer opinions on the future tendency of inflation (\texttt{CIE}). The baseline set of variables also includes our measure of uncertainty (see below) as endogenous variable. Details on data sources and a priori transformations are provided in Appendix \ref{app:C}. The number of endogenous variables is $M=13$. We choose a lag length of $P=4$.

As our measure for EA wide uncertainty in $u_t$, we use the economic policy uncertainty (\texttt{EPU}) index developed by \citet{baker2016measuring} on the log-scale.\footnote{The series is available for download from \href{http://www.policyuncertainty.com}{policyuncertainty.com}.} This newspaper-based index is chosen on the grounds of it capturing a broad definition of uncertainty (as opposed to strictly focusing on macroeconomic or financial uncertainty). A comparison of uncertainty indices for the EA and its countries is provided in \citet{rossi2017macroeconomic} or \citet{azqueta2020economic}. Since the index shows upward trending behavior in its raw format, we detrend and standardize it prior to including it in our model (that is, the resulting measure has zero mean and unit variance). We achieve this by regressing the series on a linear trend term, and subtract the fitted values from this regression from the original series. When used without these adjustments, the index essentially splits our sample into pre- and post-sovereign debt crisis regimes. 

Our detrending procedure allows to capture deviations of uncertainty from a hypothetical long-run equilibrium, thereby accounting for the fact that economic agents may react endogenously to prolonged periods with high-levels of uncertainty and specific uncertainty-related events. For related studies on endogenous movements of uncertainty, see \citet{carriero2018endogenous} and \citet{ludvigson2020uncertainty}. By contrast, the detrended series marks increases/decreases of uncertainty relative to a baseline level, providing a more adequate picture of uncertainty episodes.

\section{Empirical results}\label{sec:results}
\subsection{Uncertainty indicator and state allocation}
\begin{figure}[t]
\includegraphics[width=\textwidth]{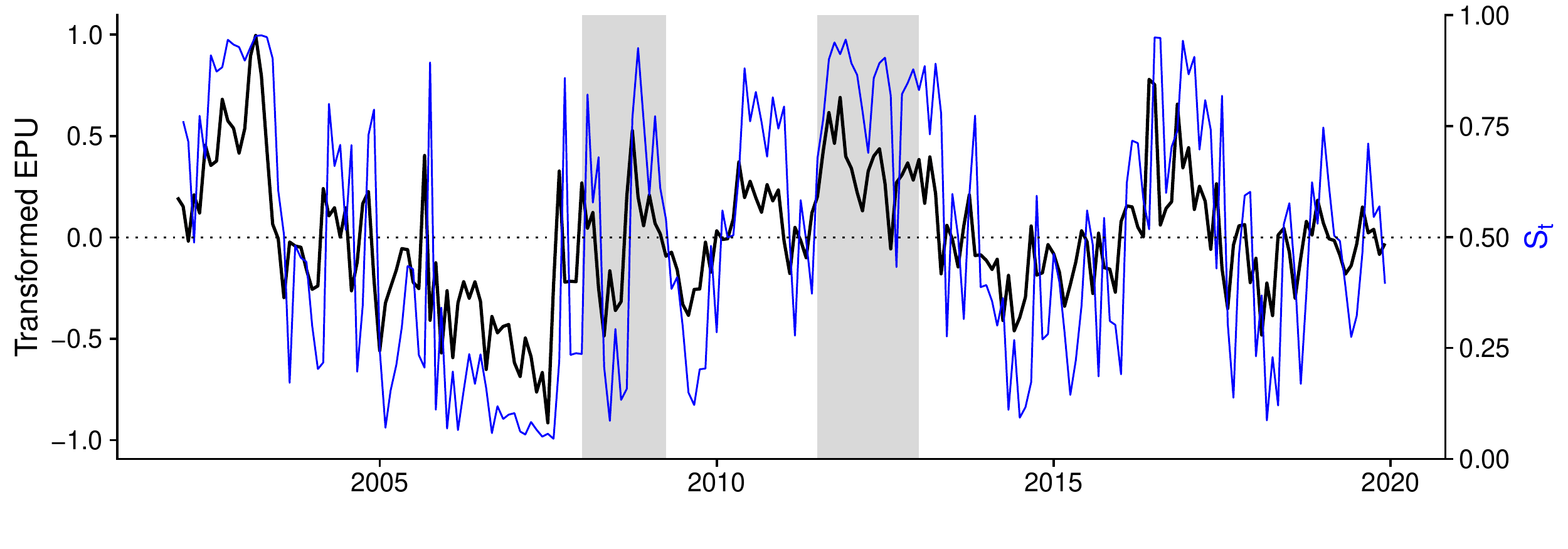}
\caption{Transformed uncertainty measure $u_t$ and the posterior median of state allocation $S_t$.}\label{fig:uncertainty}\vspace*{-0.3cm}
\caption*{\footnotesize\textit{Notes}: The left axis shows our detrended and demeaned measure of uncertainty (economic policy uncertainty, EPU) on the log-scale (left axis, in solid black). The right axis refers to the state allocation $S_t\in[0,1]$ over time (right axis, in solid blue). Note that the posterior median of $S_t(u_{t-1})$ closely mirrors the series implied by the expected values of our priors on $\gamma$ and $\phi$, which is due to the previously discussed informativeness of our prior setup. The dotted horizontal line marks zero. The grey shaded areas indicate recessions dated by the CEPR Euro Area Business Cycle Dating Committee.}
\end{figure}

Figure \ref{fig:uncertainty} shows the evolution of the transformed measure of uncertainty (solid black, left axis). The right axis refers to the state allocation $S_t(u_{t-1})\in[0,1]$ over time (right axis, in solid blue). For this plot, we show the posterior median state allocation of $S_t(u_t{-1})$. Note that the posterior distribution closely mirrors the series implied by the expected values of our priors on $\gamma$ and $\phi$, which is due to the previously discussed informativeness of our setup.

We start with discussion particulars of the chosen uncertainty index. Highest values are observed early in the sample, during the second Iraq War starting in $2003$. From $2005$ until the onset of the Great Recession, a period of relatively low uncertainty emerges. Just before and during the Great Recession we observe several peaks in economic policy uncertainty, related to several fiscal and monetary policy measures intended to calm financial markets and foster economic recovery. Between the two major recessions of our sample, the index first lowers around $2009$, but suddenly increases again in early $2010$. The elevated levels in $2010$ mark substantial increases in sovereign credit risk throughout Europe. These dynamics culminate during the most severe years of the European debt crisis between $2012$ and $2013$. While we observe a brief period of lower uncertainty after $2014$, the index again increases around the Brexit referendum in mid-$2016$ in the United Kingdom. After $2017$, the index fluctuates around its unconditional mean.

Turning to $S_t(u_{t-1})$, we find that the main features of the uncertainty index translate closely to our bounded indicator. It is worth mentioning that several periods appear to be more persistent when compared to the measure of uncertainty. While peaks and troughs in the EPU index often yield values of $S_t(u_{t-1})$ close to one and zero respectively, a substantial period is characterized by values in between. Using the bounded indicator rather than interacting the endogenous variables of the VAR directly with uncertainty allows for estimating clearly separated low/high uncertainty regimes. While this may also be achieved via MS models, our approach has the advantage that we also take into account ``normal'' times where we neither observe particularly high or low levels of uncertainty. We thereby combine the two advantages of both interacted and MS-VAR models in our framework. This implies that conditional on the respective level of uncertainty, we can estimate time-varying effects of CMP and UMP.

\subsection{Conventional and unconventional monetary policy under uncertainty}
Our econometric framework allows to calculate impulse response functions at each point in time. This implies that both impact reactions and transmission dynamics governed by the transitioning VAR coefficients may differ over time, conditional on the uncertainty indicator. It is worth mentioning that \citet{altavilla2019measuring} find that EA financial market participants do not perceive monetary policy effects to be asymmetric regarding positive/negative surprises in terms of asset prices responses. This implies that our regime-allocation and responses will not be driven by monetary policy shocks being predominantly positive/negative in times of varying levels of uncertainty.

To obtain relative impulse responses to the shocks indexed $s\in\{\text{TG},\text{FG},\text{QE}\}$, we let $\varsigma_s$ denote the standard error of $x_{st}$. Moreover, let $\bm{M}_i$ denote the $MP\times MP$ companion matrix for regime $i\in\{0,1\}$ based on the VAR in Eq. (\ref{eq:benchmarkVAR}),\footnote{That is, the first $M$ rows of this matrix are the VAR coefficients in $\bm{A}_i$ and the remaining rows yield an identity.} and define the companion-form impact vector $\tilde{\bm{\delta}}_{is} = (\varsigma_s\bm{\delta}_{is}',\bm{0}_{1\times [(M-1)P]})'$. By multiplying $\bm{\delta}_{is}$ with $\varsigma_s$ we obtain responses reflecting a one-standard deviation shock to the instrument. The impulse response function for shock $s$ at horizon $h$ (for $h>0$) at time $t$ is:
\begin{equation*}
    \text{\textbf{IRF}}_{sh,t} = \bm{J}\left(\bm{M}_{1} \times S_t(u_{t-1}) + \bm{M}_{0} \times [1-S_t(u_{t-1})]\right)^h \tilde{\bm{\delta}}_{is},
\end{equation*}
where $\bm{J}$ is an $M\times MP$-dimensional matrix that selects the first $M$ rows of a vector of dimension $MP\times 1$. In the following figures, we distinguish shocks by the respective colors of the impulse response functions. TG shocks are in shades of blue, FG shocks in red and QE shocks in green.

The impulse response functions at time $t$ depend on the lagged uncertainty indicator $u_{t-1}$. Consequently, all subsequent plots over time are to be understood in conjunction with Fig. \ref{fig:uncertainty}. The respective time-axis (x-axis) shows the period for which set of VAR coefficients depending on the indicator $S_t(u_{t-1})$ the impulses were computed. The columns (a) to (e) of the individual figures refer to the value of the impulse response function at the specified horizon $h$ over time. The rows show variable types, with the scale of the response being determined by the transformations of the underlying data stated in Appendix \ref{app:C}.\footnote{While our econometric framework in principle allows to calculate impulse responses conditional on the state indicator $S_t = 1$ and $S_t = 0$ that correspond to pure regimes, we refrain from including them to economize on space. While weighted averages of the coefficients are for the most part stable in the sense that the maximum absolute eigenvalue of the companion matrix is below one and the system returns to its equilibrium at higher-order horizons, this is not necessarily the case for the hypothetical cleanly separated regimes. Relatedly, a major goal of this article is to illuminate the effectiveness of monetary policy during specific economic circumstances, which we achieve by showing impulse response functions over time conditional on the actual prevailing level of uncertainty. Additional results on the responses for cleanly separated regimes are available from the corresponding author upon request.}

\subsubsection{The target/policy rate shock}
The first set of results shows the responses to a CMP shock, captured by the target factor. We split our discussion into the three variable blocks relating to financial (Fig. \ref{fig:irfs_TG_fin}), macroeconomic (Fig. \ref{fig:irfs_TG_macro}) and survey-based (Fig. \ref{fig:irfs_TG_survey}) quantities. These figures (as do those for the FG and QE shock) collect the responses of variables in the respective rows, while the columns refer to the horizon $h$.

To assess direct effects of monetary policy shocks on financial quantities, consider Fig. \ref{fig:irfs_TG_fin}. Recall that the responses are scaled to reflect a one-standard deviation shock to the instrument. A contractionary CMP shock results in a contemporaneous increase of E3M, which is targeted by the policy rates, around one basis point. It is worth mentioning that this increase stays roughly the same with respect to the time-axis, indicating that uncertainty has negligible effects on the impact response (apart from lower values of uncertainty widening the posterior credible set). 

\setcounter{figure}{0}
\renewcommand{\thefigure}{3.\arabic{figure}}
\begin{figure}[ht]
    \captionsetup[subfigure]{justification=centering}
    \begin{subfigure}[t]{0.19\textwidth}
        \includegraphics[width=\textwidth]{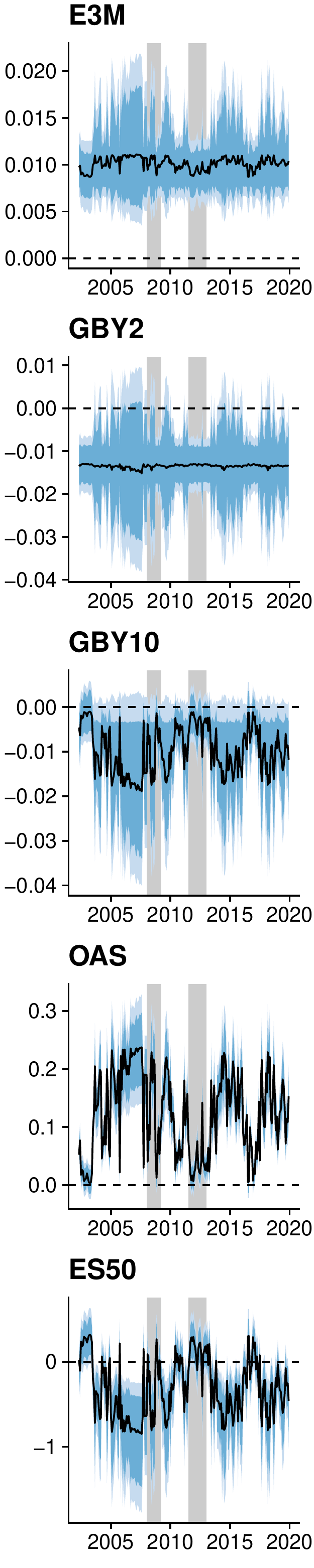}
        \caption{$h=0$}
    \end{subfigure}
    \begin{subfigure}[t]{0.19\textwidth}
        \includegraphics[width=\textwidth]{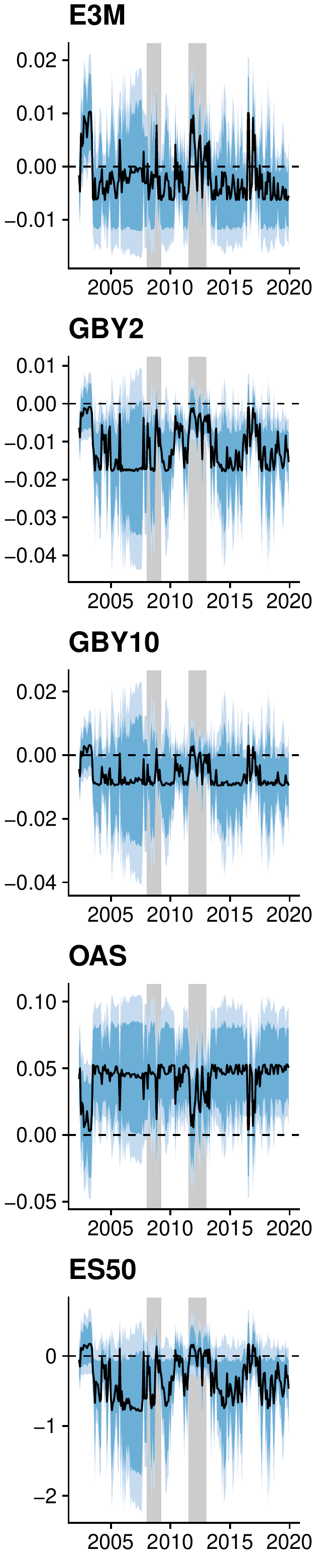}
        \caption{$h=4$}
    \end{subfigure}
    \begin{subfigure}[t]{0.19\textwidth}
        \includegraphics[width=\textwidth]{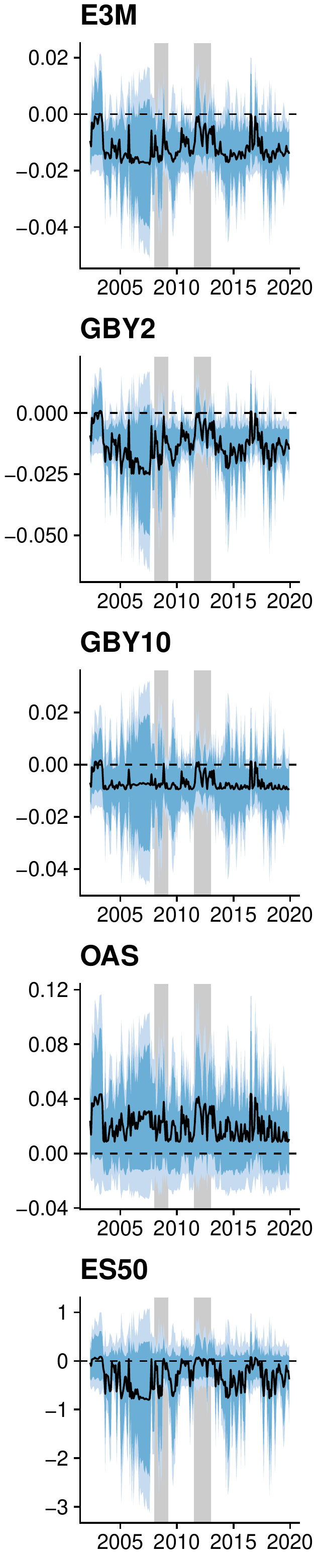}
        \caption{$h=12$}
    \end{subfigure}
    \begin{subfigure}[t]{0.19\textwidth}
        \includegraphics[width=\textwidth]{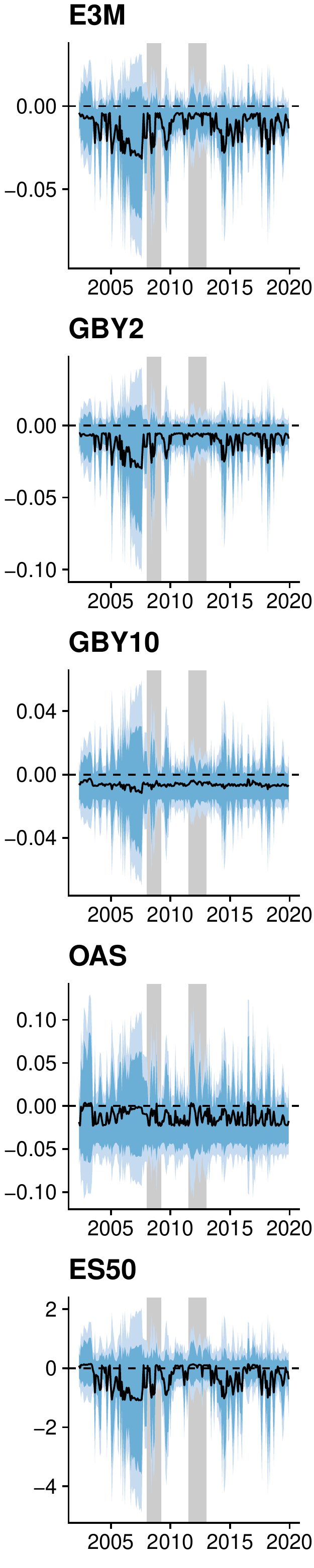}
        \caption{$h=24$}
    \end{subfigure}
    \begin{subfigure}[t]{0.19\textwidth}
        \includegraphics[width=\textwidth]{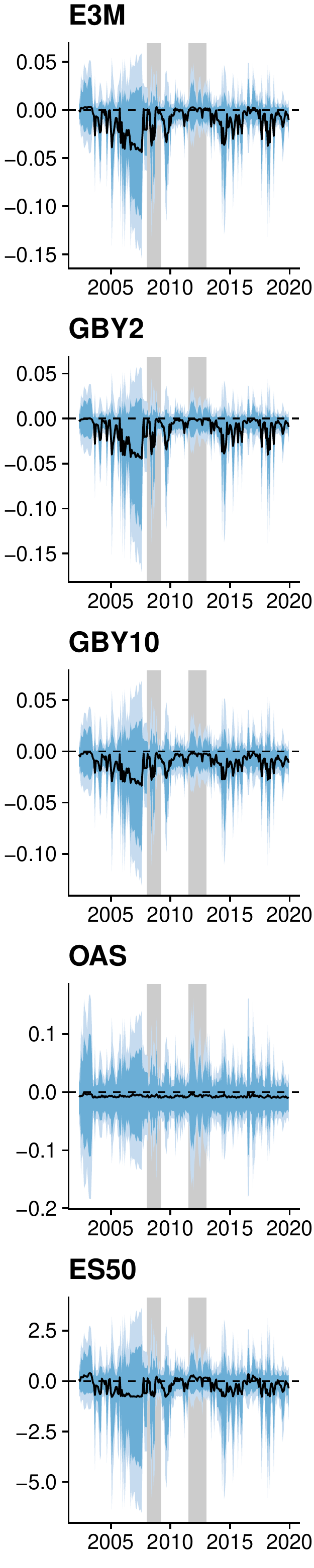}
        \caption{$h=36$}
    \end{subfigure}
    \caption{Responses of financial variables to the TG shock at different horizons.}\label{fig:irfs_TG_fin}\vspace*{-0.3cm}
    \caption*{\footnotesize\textit{Notes}: The sample runs from 2002:01 to 2019:12. The horizontal dashed black line marks zero. The solid black line is the posterior median of the impulse response function for period $t$ at different horizons $h=\{0,4,12,24,36\}$, that is, the impact response, the one-quarter, one-year, two-years and three-years ahead response. The dark shades indicate the $50$ percent posterior credible set, the lighter shade the $68$ percent posterior credible set. The grey shaded areas indicate recessions dated by the CEPR Euro Area Business Cycle Dating Committee. For variable codes see Section \ref{sec:data} and Appendix \ref{app:C}.}
\end{figure}

Two-year yields show simultaneous decreases of about 1.5 basis points, with similar persistence over time. This differs with respect to ten-year government bond yields, where we detect substantial variation over time. In particular, we find that negative impact reactions for longer-maturity yields are stronger during periods of low uncertainty. Relevant periods in the sample are, for instance, from 2005 until the financial crisis in 2007/08, and to a lesser extent, low-uncertainty periods after the sovereign debt crisis until the Brexit referendum in 2016. On the contrary, effects are much smaller during high-uncertainty periods. 

Turning to OAS, our measure capturing the tightness of financial conditions, we observe several interesting findings. While the impulse for this variable is significantly positive for periods of low uncertainty, it is muted and even insignificant during recessionary episodes (e.g., during the financial crisis and European debt crisis, but also early in the sample during the second Iraq War). While contractionary shocks increase spreads (which is in line with theory), this channel appears to be impaired under uncertain economic conditions. Our final variable is the Euro Stoxx 50 index, where the scale is in percent. Again, we detect substantial time-variation in the impact response. Particularly during high-uncertainty periods, where spreads also show muted reactions, we detect positive effects of contractionary policy on stock prices.\footnote{The stock market response to monetary policy shocks is typically less pronounced for EA data when compared to the US, see \citet{jarocinski2020deconstructing}.} In comparatively certain periods, we observe effects of about one percent declines that are in line with the previous literature. While this may seem puzzling at first, we conjecture that this finding relates to information effects occurring particularly under uncertainty. As suggested by \citet{jarocinski2020deconstructing}, information shocks may reverse the responses of CMP shocks, thereby muting or overruling overall effects. We argue that this is one of the reasons why conventional monetary policy appears to be less effective when uncertainty is elevated.

Turning to higher order effects, we find that time-varying patterns measured in the impacts do not necessarily carry over to responses at higher horizons. For E3M, for example, we detect substantial time-variation for one-quarter ahead estimates. In periods of high uncertainty, our findings imply that shocks are more persistent, and we find that interest rates turn negative briefly around the one-year horizon. For two-year and ten-year yields, the opposite is the case, with shocks neutralizing quicker during high-uncertainty periods. For our spread variable and the stock market, we detect decreasing importance of time-variation with respect to the impulse response horizon. For OAS, this implies that spreads react more sluggish under uncertainty, and the shock peters out around the one-year ahead horizon.

Following our discussion of direct effects measured with financial variables, we proceed with investigating the effects on several key macroeconomic quantities in Fig. \ref{fig:irfs_TG_macro}. The impaired transmission of CMP shocks on financial quantities carries over to these indicators. While all our responses are in line with economic theory, the magnitude of the responses differs substantially over time.

\begin{figure}[ht]
    \captionsetup[subfigure]{justification=centering}
    \begin{subfigure}[t]{0.19\textwidth}
        \includegraphics[width=\textwidth]{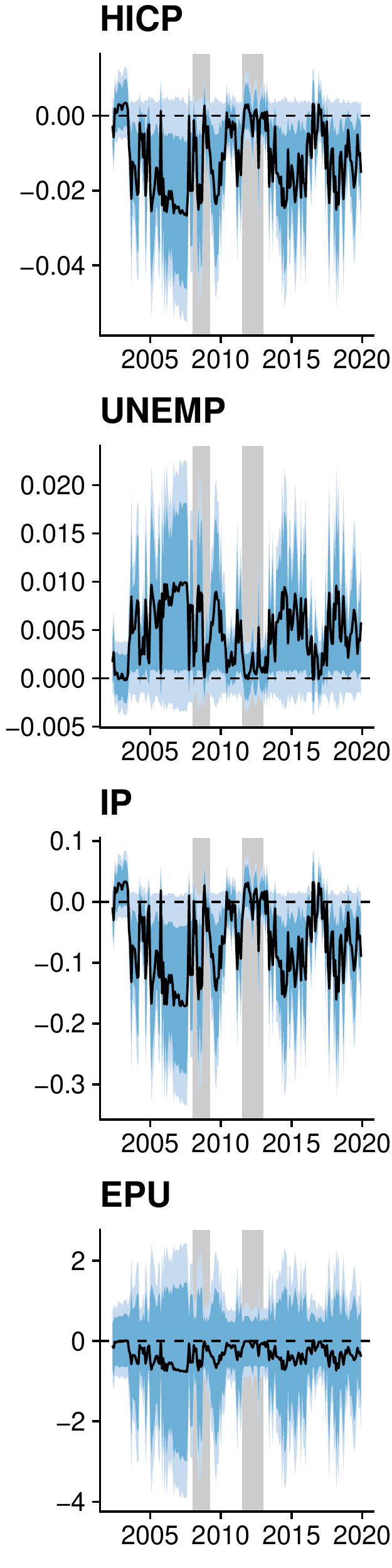}
        \caption{$h=0$}
    \end{subfigure}
    \begin{subfigure}[t]{0.19\textwidth}
        \includegraphics[width=\textwidth]{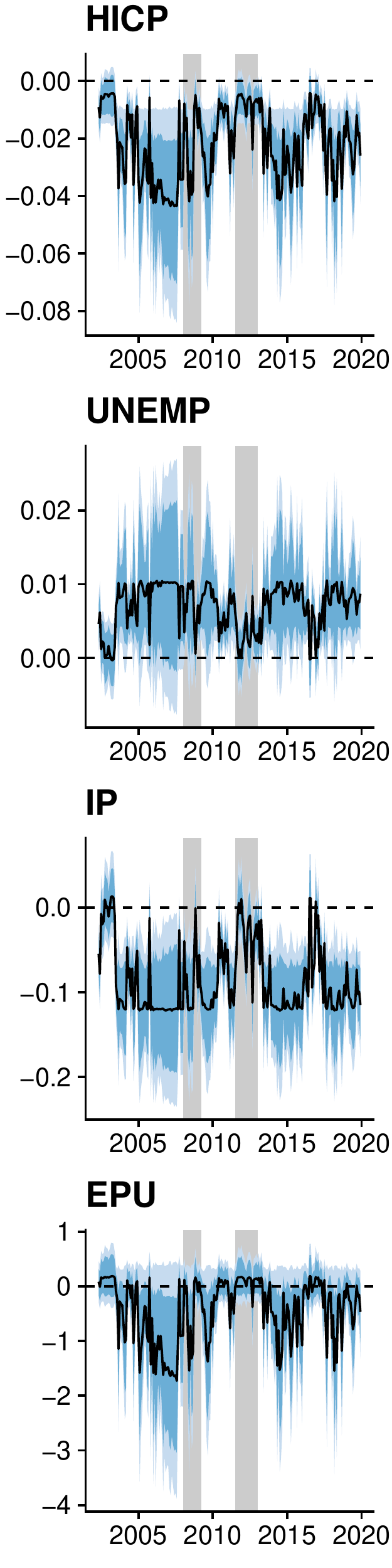}
        \caption{$h=4$}
    \end{subfigure}
    \begin{subfigure}[t]{0.19\textwidth}
        \includegraphics[width=\textwidth]{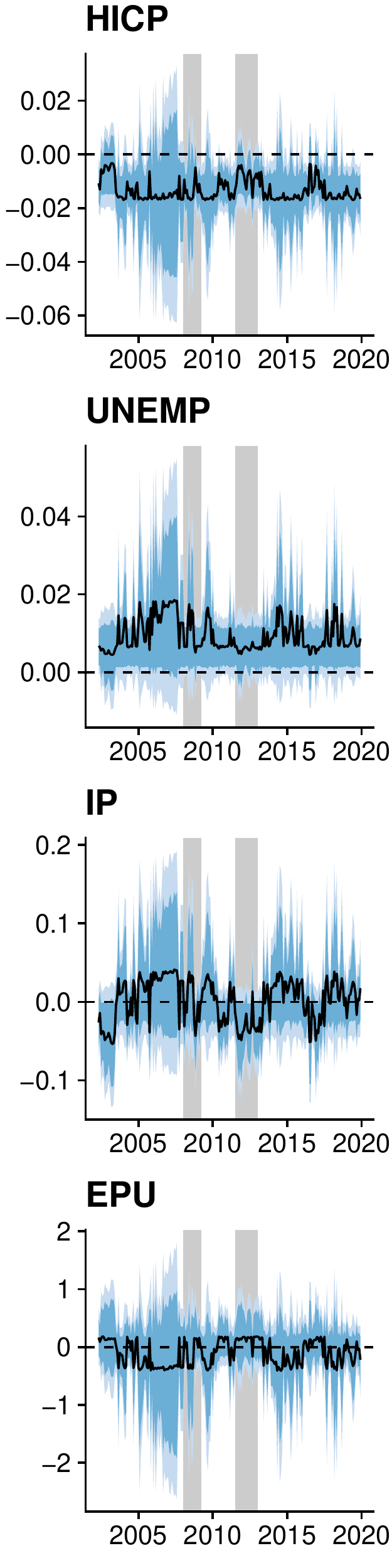}
        \caption{$h=12$}
    \end{subfigure}
    \begin{subfigure}[t]{0.19\textwidth}
        \includegraphics[width=\textwidth]{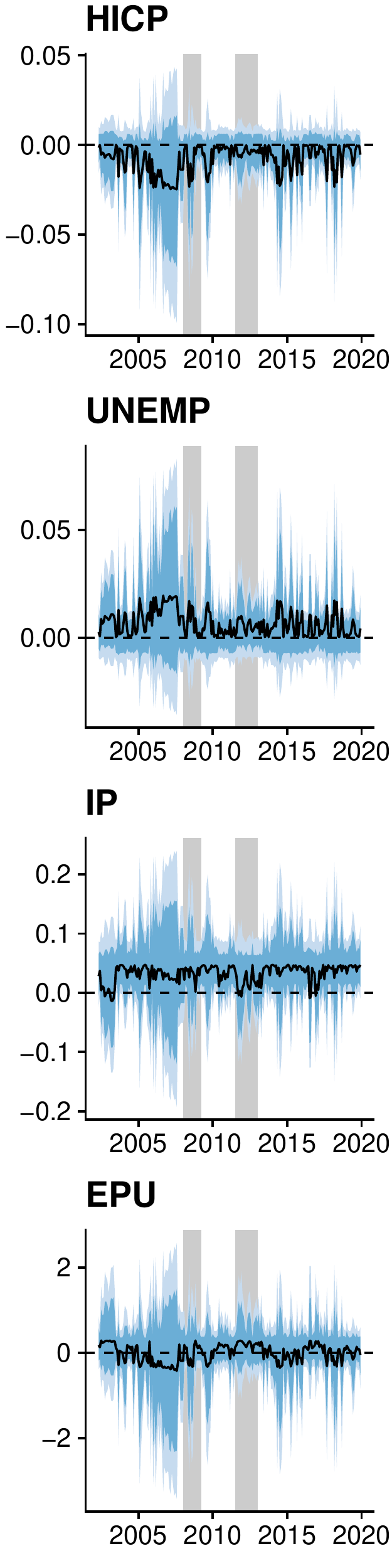}
        \caption{$h=24$}
    \end{subfigure}
    \begin{subfigure}[t]{0.19\textwidth}
        \includegraphics[width=\textwidth]{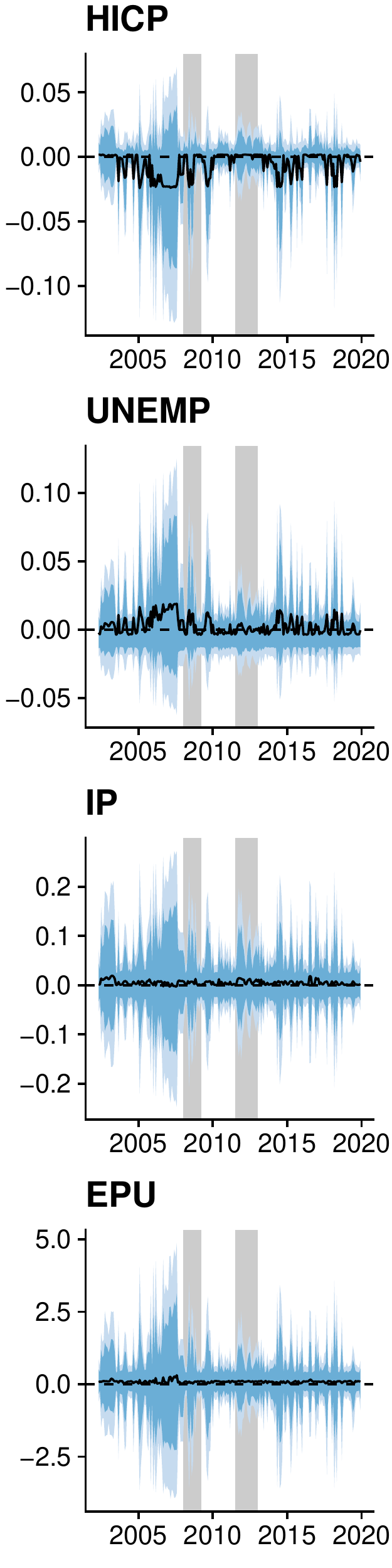}
        \caption{$h=36$}
    \end{subfigure}
    \caption{Responses of macroeconomic variables to the TG shock at different horizons.}\label{fig:irfs_TG_macro}\vspace*{-0.3cm}
    \caption*{\footnotesize\textit{Notes}: The sample runs from 2002:01 to 2019:12. The horizontal dashed black line marks zero. The solid black line is the posterior median of the impulse response function for period $t$ at different horizons $h=\{0,4,12,24,36\}$, that is, the impact response, the one-quarter, one-year, two-years and three-years ahead response. The dark shades indicate the $50$ percent posterior credible set, the lighter shade the $68$ percent posterior credible set. The grey shaded areas indicate recessions dated by the CEPR Euro Area Business Cycle Dating Committee. For variable codes see Section \ref{sec:data} and Appendix \ref{app:C}.}
\end{figure}

We find that inflation and economic activity (measured by industrial production) decline on impact after a contractionary shock during low-uncertainty episodes as expected. However, when uncertainty is high, we often measure insignificant effects. Especially informative in this context are the two recessions in our sample. The same is true for the unemployment rate, albeit with reversed sign. Contractionary shocks usually translate to increases in unemployment, which is exactly what we find for cases where $S_t(u_{t-1})$ is close to zero. When uncertainty is elevated, CMP appears to be less effective, and we find that unemployment does not react to manipulations of the target rate significantly and with substantial delay. The uncertainty measure, on the other hand, does not react contemporaneously to CMP in a significant manner regardless of the prevailing level of uncertainty.

Higher order effects show substantial persistence in our responses. Expected effects (lower inflation and industrial production, higher unemployment rate), albeit with small magnitudes, are estimated significantly for the most part at the one-year ahead horizon. It is also interesting to note that our uncertainty measure appears to respond endogenously to monetary policy measures at the one-quarter ahead horizon, with large effects during low-uncertainty periods of about two percent decreases in uncertainty. At the two-year ahead impulse response horizon, our posterior credible sets indicate that the shocks are neutralized and the system returns to its equilibrium.

 These results are in line with \citet{aastveitetal2017} who evaluate US monetary policy shocks in times of high and low uncertainty. Within their framework, they also find that effects of monetary policy tend to be much smaller when uncertainty is high. Indeed, they find that monetary policy effects on real economic activity are approximately halved when their measure of uncertainty is in its upper decile as compared to it being in this bottom decile, and the difference is statistically significant. Similarly, \citet{caggianoetal2017} find that the stabilizing effects of monetary policy after an uncertainty shock seem to be greater in times of expansion, rather than contraction. This indicates that our results on the effects of CMP given different levels of uncertainty are in line with the existing literature on the US.

\begin{figure}[ht]
    \captionsetup[subfigure]{justification=centering}
    \begin{subfigure}[t]{0.19\textwidth}
        \includegraphics[width=\textwidth]{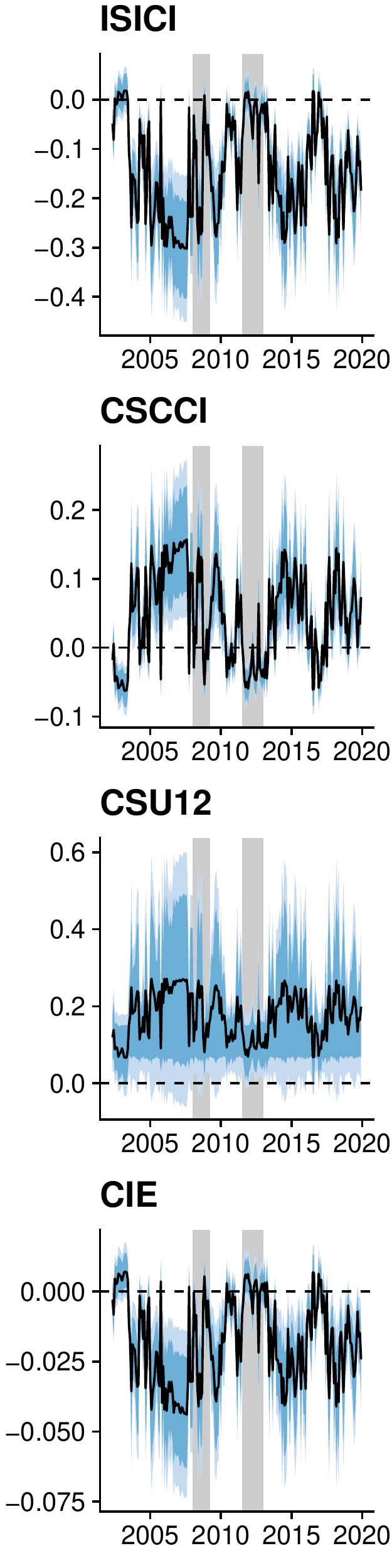}
        \caption{$h=0$}
    \end{subfigure}
    \begin{subfigure}[t]{0.19\textwidth}
        \includegraphics[width=\textwidth]{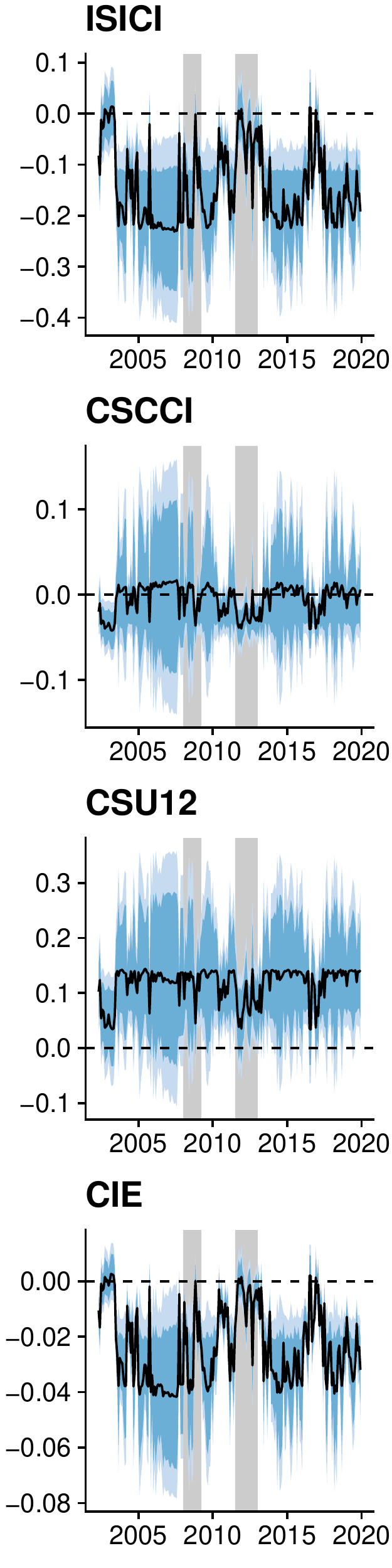}
        \caption{$h=4$}
    \end{subfigure}
    \begin{subfigure}[t]{0.19\textwidth}
        \includegraphics[width=\textwidth]{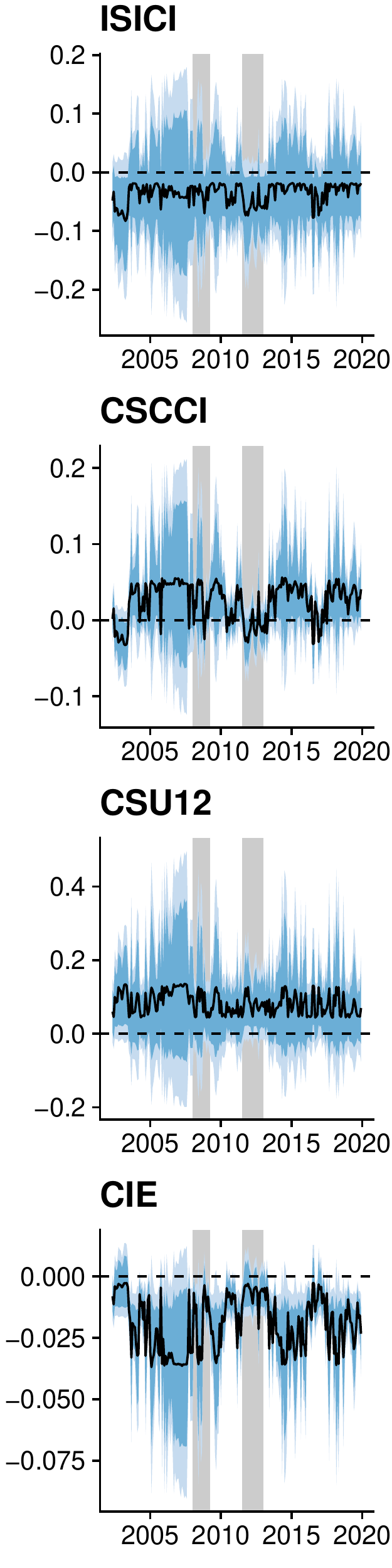}
        \caption{$h=12$}
    \end{subfigure}
    \begin{subfigure}[t]{0.19\textwidth}
        \includegraphics[width=\textwidth]{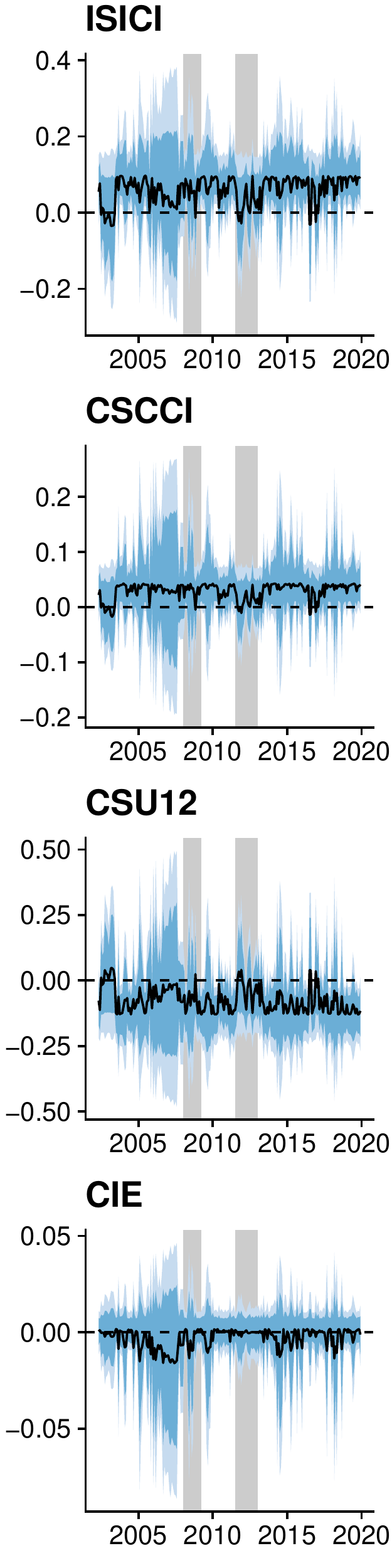}
        \caption{$h=24$}
    \end{subfigure}
    \begin{subfigure}[t]{0.19\textwidth}
        \includegraphics[width=\textwidth]{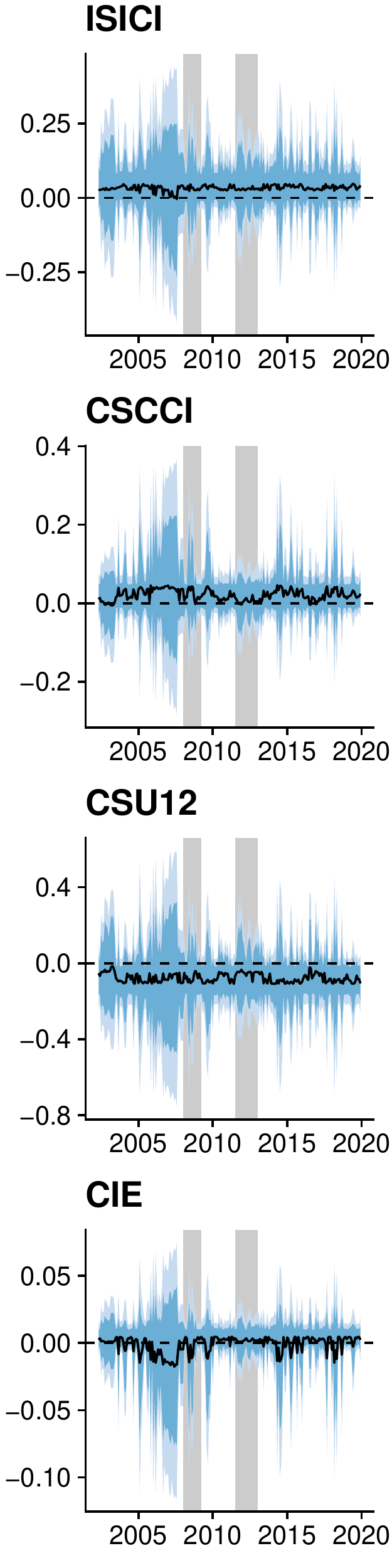}
        \caption{$h=36$}
    \end{subfigure}
    \caption{Responses of survey-based expectations to the TG shock at different horizons.}\label{fig:irfs_TG_survey}\vspace*{-0.3cm}
    \caption*{\footnotesize\textit{Notes}: The sample runs from 2002:01 to 2019:12. The horizontal dashed black line marks zero. The solid black line is the posterior median of the impulse response function for period $t$ at different horizons $h=\{0,4,12,24,36\}$, that is, the impact response, the one-quarter, one-year, two-years and three-years ahead response. The dark shades indicate the $50$ percent posterior credible set, the lighter shade the $68$ percent posterior credible set. The grey shaded areas indicate recessions dated by the CEPR Euro Area Business Cycle Dating Committee. For variable codes see Section \ref{sec:data} and Appendix \ref{app:C}.}
\end{figure}

Our final set of results for survey-based expectations in response to the TG shock are displayed in Fig. \ref{fig:irfs_TG_survey}. We detect substantial time-variation in impact responses conditional on the uncertainty indicator. In particular, responses for most variables are again muted under uncertainty, evidenced by a brief high-uncertainty period early in the sample, the two recessions and around the Brexit referendum in 2016. This points towards impaired transmission channels of monetary policy during these episodes. Industrial, unemployment and inflation expectations show insignificant responses during these periods where $S_t(u_{t-1})$ is close or equal to one, while impacts are often substantial for the indicator when it is close to zero. This finding may be linked to \citet{coibion2018firms}, who find that expectation formation in firms is often based on suboptimal information. An argument in this context is that it is even harder to update expectations in light of high-levels of uncertainty. 

A puzzling result is apparent for consumer expectations. Here, we find that during uncertain episodes, confidence decreases in response to contractionary shocks as expected. During more certain times, however, consumer expectations appear to improve after such a shock. We find different levels of persistence with respect to individual variables, with particularly persistent effects of inflation expectations. This relates to the survey-based findings for a randomized-controlled trial by \citet{coibion2019monetary}, who identify substantial cross-sectional dispersion of inflation expectations. This suggests that while some consumers can correctly identify monetary policy actions, others cannot, and this translates to persistent effects of the shocks. We conjecture that high uncertainty hinders updating expectations across subsets of informed and uninformed consumers disproportionately, resulting in particularly less effective transmission in uncertain times.

Summarising, we find that while a contractionary CMP shocks result in comparable increases in interest rates independently from the level of uncertainty, this translates into tighter financial conditions more efficiently in times of low uncertainty. Our financial market indicator Euro Stoxx 50 index even shows positive effects in times of elevated uncertainty. These results carry over to indicators of the real economy. Episodes of low uncertainty show textbook reactions of real variables like inflation, economic activity and unemployment to a contractionary CMP shock. In contrast, in times of higher uncertainty, we find insignificant or delayed effects. As for our variables of survey-based expectations, again, CMP seems to affect expectations less in times of high uncertainty. All these findings point to the conclusion that CMP is a relevant and efficient tool in times of low or medium uncertainty. However, its transmission to financial markets, real economic activity and expectations is impaired when uncertainty is higher. 

It is possible that transmission via investment or consumption based channels, such as the direct interest rate channel, the intertemporal substitution effect or wealth effects \citep{boivinmishkin2010, mishkin1996}, might work differently or less effectively when uncertainty is high. We conjecture that one reason for this impairment might be that consumers do not change their inflation expectations after a CMP shock hits the economy, a result that we find in our model. As it is the real interest rate rather than the nominal interest rate that affects asset prices and spending, it is possible that changed dynamics in the formation of inflation expectations in times of high uncertainty also lead to different or even diverging dynamics between real and nominal interest rates, resulting in hindered transmission of CMP. Overall, all these findings are in line with existing literature that suggest that real option effects from theory may exist and that elevated uncertainty leads economic agents to form expectations not rationally or perfectly informed, and perhaps, even to postpone decision making until uncertainty decreases.

\subsubsection{The forward guidance shock}
Turning to the first measure capturing UMP, we discuss our results for the FG shock based on the results displayed in Figs. \ref{fig:irfs_FG_fin}, \ref{fig:irfs_FG_macro} and \ref{fig:irfs_FG_survey}. Again, we simulate the response of the endogenous variables to a contractionary one-standard deviation shock to the instrument.

\setcounter{figure}{0}
\renewcommand{\thefigure}{4.\arabic{figure}}
\begin{figure}[ht]
    \captionsetup[subfigure]{justification=centering}
    \begin{subfigure}[t]{0.19\textwidth}
        \includegraphics[width=\textwidth]{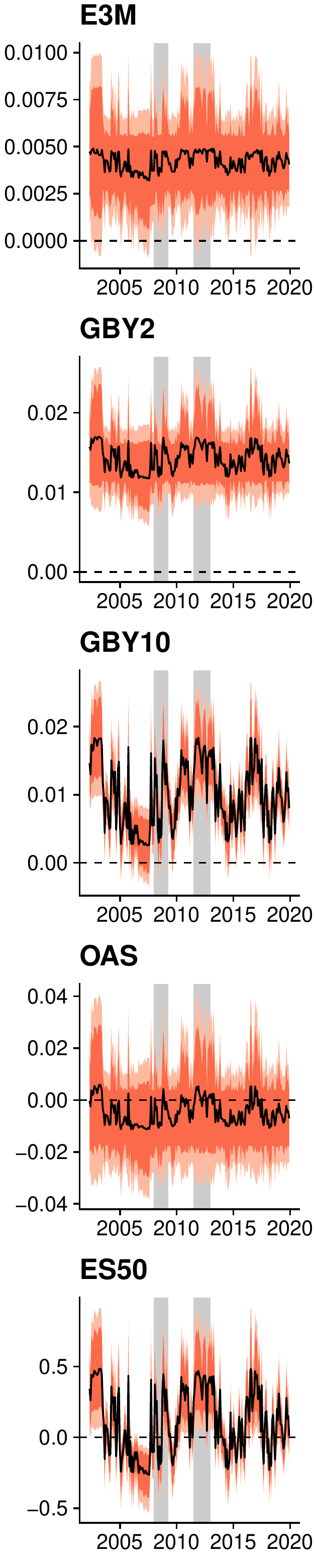}
        \caption{$h=0$}
    \end{subfigure}
    \begin{subfigure}[t]{0.19\textwidth}
        \includegraphics[width=\textwidth]{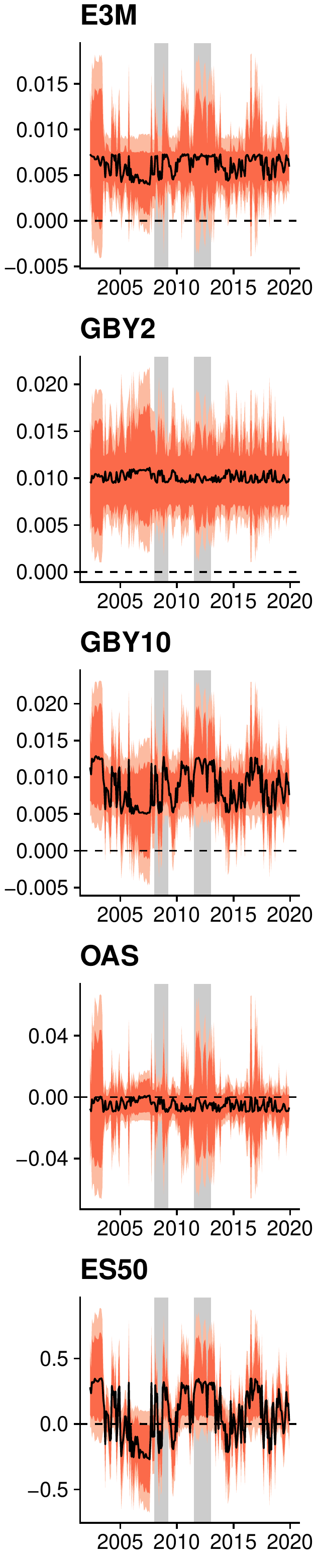}
        \caption{$h=4$}
    \end{subfigure}
    \begin{subfigure}[t]{0.19\textwidth}
        \includegraphics[width=\textwidth]{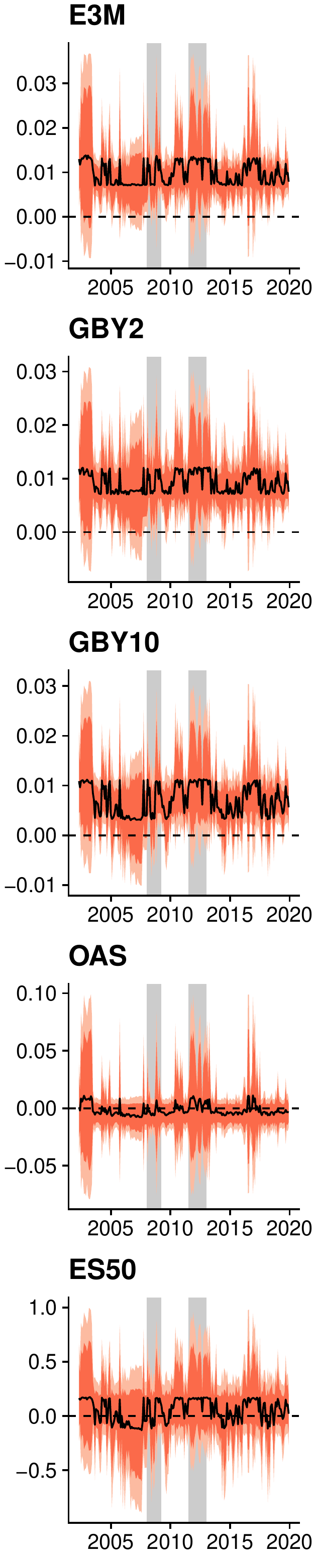}
        \caption{$h=12$}
    \end{subfigure}
    \begin{subfigure}[t]{0.19\textwidth}
        \includegraphics[width=\textwidth]{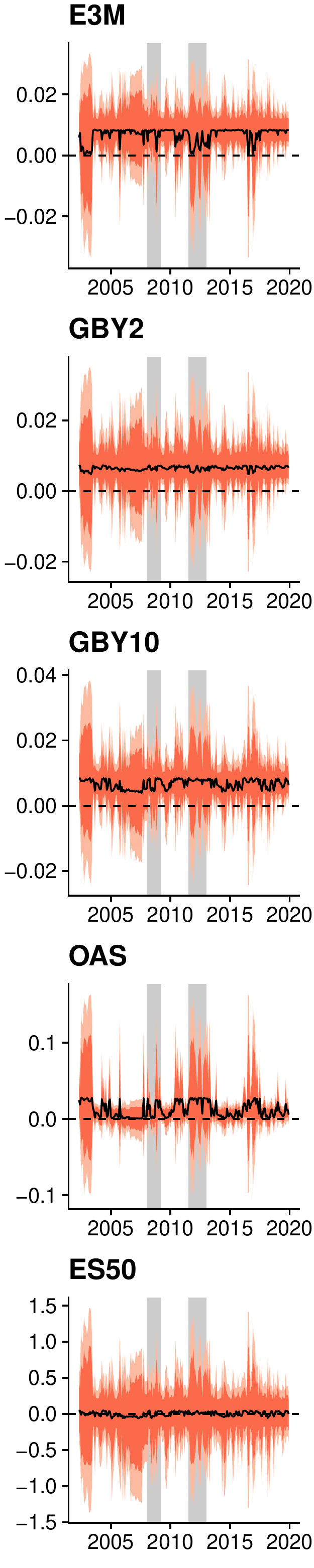}
        \caption{$h=24$}
    \end{subfigure}
    \begin{subfigure}[t]{0.19\textwidth}
        \includegraphics[width=\textwidth]{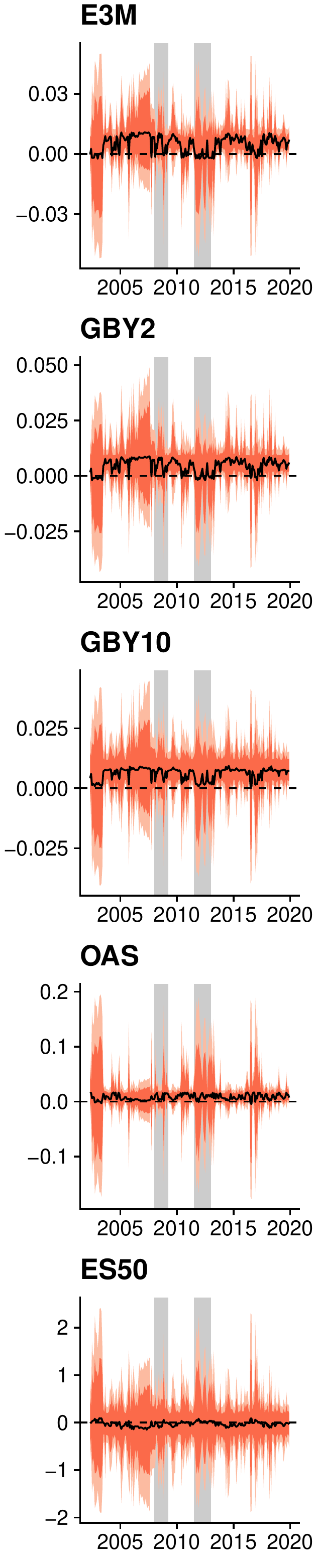}
        \caption{$h=36$}
    \end{subfigure}
    \caption{Responses of financial variables to the FG shock at different horizons.}\label{fig:irfs_FG_fin}\vspace*{-0.3cm}
    \caption*{\footnotesize\textit{Notes}: The sample runs from 2002:01 to 2019:12. The horizontal dashed black line marks zero. The solid black line is the posterior median of the impulse response function for period $t$ at different horizons $h=\{0,4,12,24,36\}$, that is, the impact response, the one-quarter, one-year, two-years and three-years ahead response. The dark shades indicate the $50$ percent posterior credible set, the lighter shade the $68$ percent posterior credible set. The grey shaded areas indicate recessions dated by the CEPR Euro Area Business Cycle Dating Committee. For variable codes see Section \ref{sec:data} and Appendix \ref{app:C}.}
\end{figure}

Figure \ref{fig:irfs_FG_fin} shows the responses of the financial variables to the FG shock. Note that in this case, the policy relevant variable are the two-year government bond yields, since the underlying factor is designed to load most strongly on this maturity. Indeed, we find the strongest reaction to the one-standard deviation of the shock at just below two basis points for the two-year government bond yield. Similar to the CMP shock discussed above, we detect only muted time-variation conditional on the prevailing level of uncertainty in the short-term rate and the two-year yield. While all three interest rates react positively on impact, the ten-year government bond yields do so more strongly in times of elevated uncertainty. Interestingly, and differently to the CMP shock, the spreads do not show significant reaction at any horizon we consider. While overall dynamics appear to be similar also for stock prices, we detect much stronger positive responses of the stock market under uncertainty. This again points to our previous notion of information effects playing a role in impairing the transmission of monetary policy during times of uncertainty.

In line with the forward looking character of this UMP instrument, we detect substantially more persistent effects of the FG shock, particularly for interest rates. Some periods, especially those characterized by low levels of uncertainty, exhibit significant positive effects even at the three-year ahead horizon. Interestingly, time-variation for higher-order impulse responses in the case of the short-term rate and two-year yields is muted, different to the CMP shock.

Figure \ref{fig:irfs_FG_macro} shows the impulses of the macroeconomic variables to the FG shock. In this context, we obtain several puzzling results. While we again find significant effects of the policy shock for inflation mainly during comparatively certain times, they exhibit the wrong sign. We conjecture that forward guidance is particularly prone to contain information effects. The responses for industrial production and unemployment have the correct sign and they are insignificant for the most part of our sample with only a muted degree of time variation. Different to CMP, we observe significant impact effects of FG on our uncertainty measure. While contractionary forward guidance lowers uncertainty in uncertain periods, it appears to have increased uncertainty during low uncertainty episodes. In terms of persistence of the shocks, we observe only a brief half-life of FG on macroeconomic quantities, with most shocks being either insignificant on impact, or petering out quickly between the one-quarter and one-year ahead horizon.

\begin{figure}[ht]
    \captionsetup[subfigure]{justification=centering}
    \begin{subfigure}[t]{0.19\textwidth}
        \includegraphics[width=\textwidth]{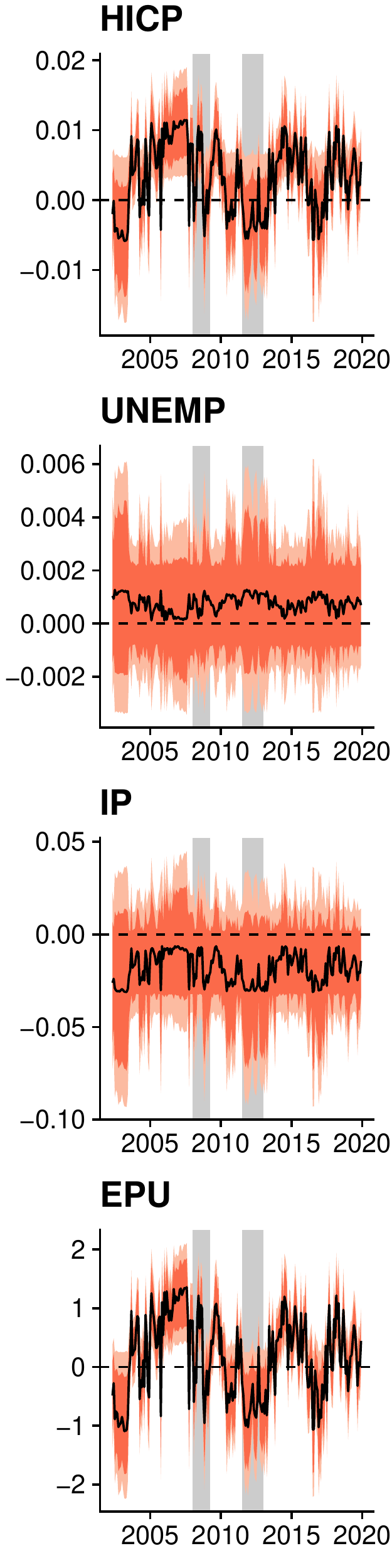}
        \caption{$h=0$}
    \end{subfigure}
    \begin{subfigure}[t]{0.19\textwidth}
        \includegraphics[width=\textwidth]{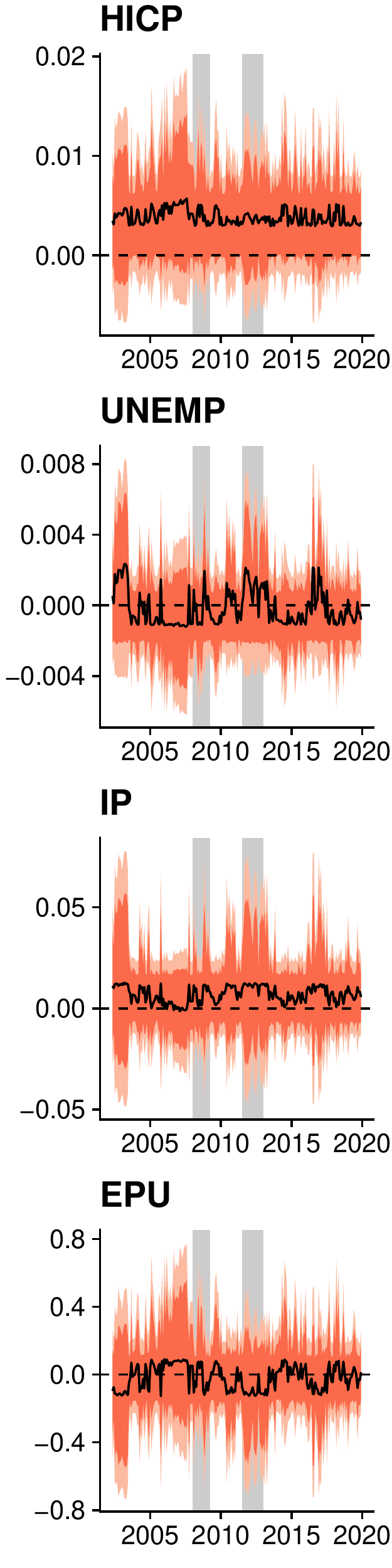}
        \caption{$h=4$}
    \end{subfigure}
    \begin{subfigure}[t]{0.19\textwidth}
        \includegraphics[width=\textwidth]{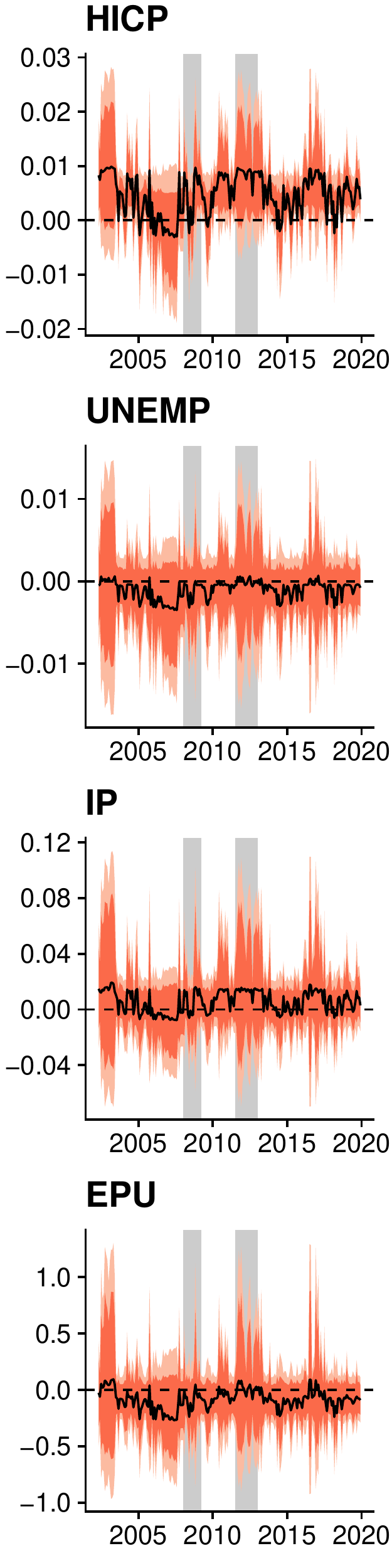}
        \caption{$h=12$}
    \end{subfigure}
    \begin{subfigure}[t]{0.19\textwidth}
        \includegraphics[width=\textwidth]{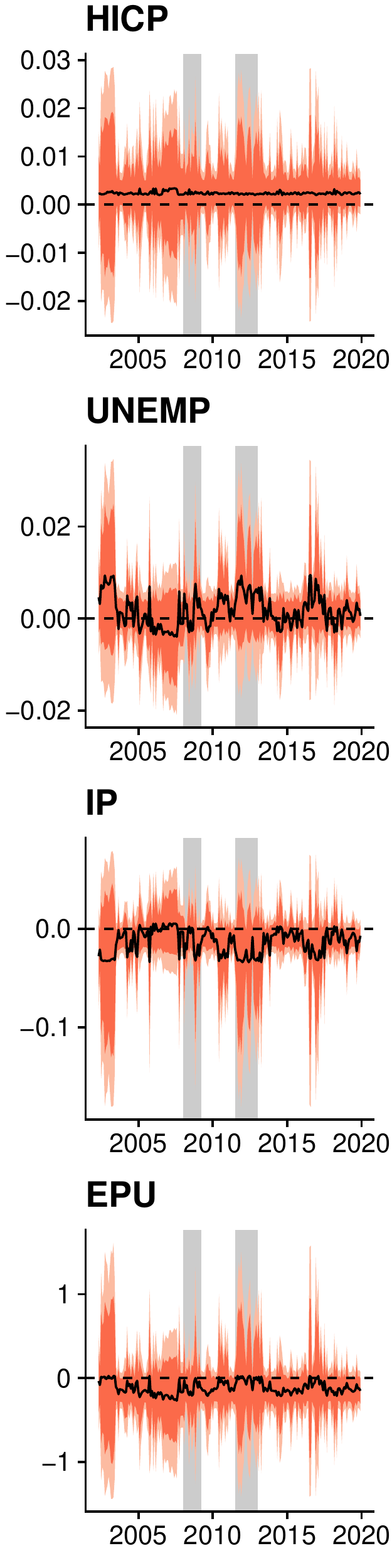}
        \caption{$h=24$}
    \end{subfigure}
    \begin{subfigure}[t]{0.19\textwidth}
        \includegraphics[width=\textwidth]{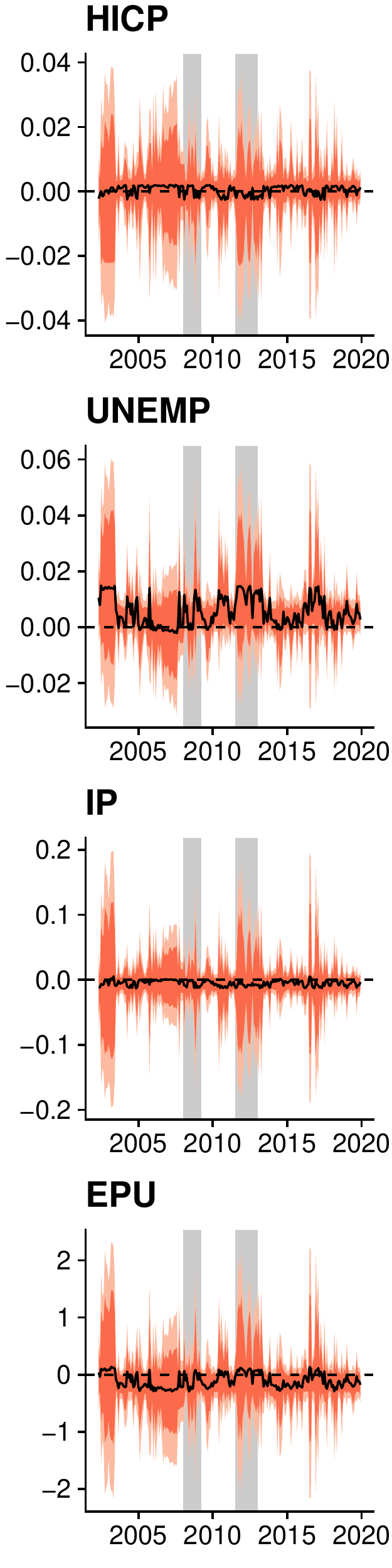}
        \caption{$h=36$}
    \end{subfigure}
    \caption{Responses of macroeconomic variables to the FG shock at different horizons.}\label{fig:irfs_FG_macro}\vspace*{-0.3cm}
    \caption*{\footnotesize\textit{Notes}: The sample runs from 2002:01 to 2019:12. The horizontal dashed black line marks zero. The solid black line is the posterior median of the impulse response function for period $t$ at different horizons $h=\{0,4,12,24,36\}$, that is, the impact response, the one-quarter, one-year, two-years and three-years ahead response. The dark shades indicate the $50$ percent posterior credible set, the lighter shade the $68$ percent posterior credible set. The grey shaded areas indicate recessions dated by the CEPR Euro Area Business Cycle Dating Committee. For variable codes see Section \ref{sec:data} and Appendix \ref{app:C}.}
\end{figure}

In our final set of results for the survey-based measures, we again detect several puzzling results. They are shown in Fig. \ref{fig:irfs_FG_survey}. We measure positive changes to industry and inflation expectations in response to a contractionary shock, albeit they are small in size and barely significant. It is worth mentioning that impact responses for industry, unemployment and inflation expectations are the same irrespective of the level of uncertainty. Consumer expectations differ substantially over time, with positive responses during elevated periods of uncertainty, and negative estimates when uncertainty is low. This finding is in line with the argument that FG conveys information about the future stance of the economy to households and consumers. The puzzling effects tend to fade away for higher-order responses, with industry, consumer and unemployment expectations exhibiting the expected sign at the two-year ahead horizon. For inflation expectations, this is not the case, and the significant positive impulse responses return to zero after two-years. Again, we note that consumer expectations are subject to substantial cross-sectional dispersion \citep{coibion2019monetary}, and this response may be an artefact of a substantial number of households being unable to update their expectations due to having difficulties identifying the specific policy action by the central bank.

\begin{figure}[ht]
    \captionsetup[subfigure]{justification=centering}
    \begin{subfigure}[t]{0.19\textwidth}
        \includegraphics[width=\textwidth]{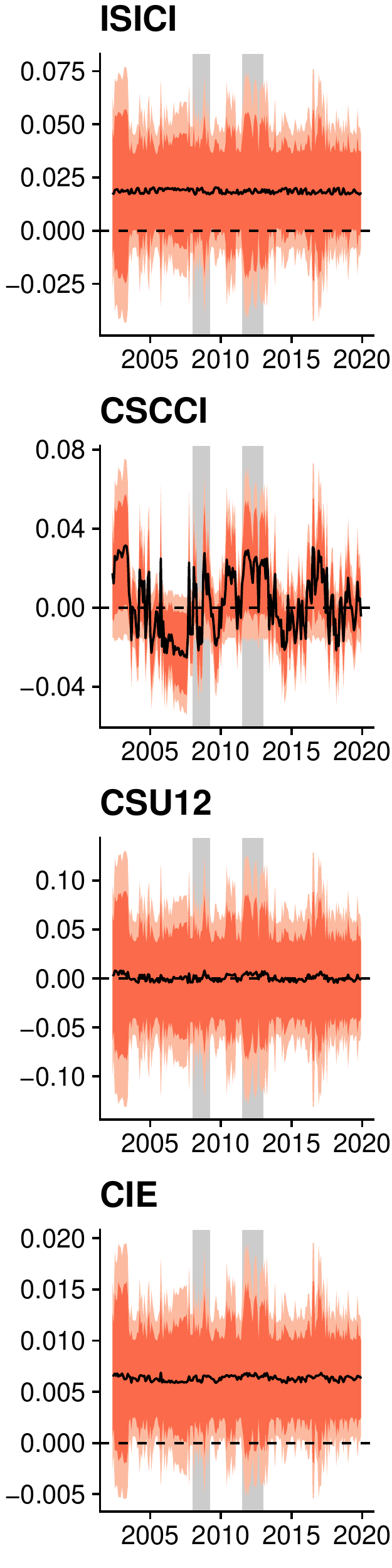}
        \caption{$h=0$}
    \end{subfigure}
    \begin{subfigure}[t]{0.19\textwidth}
        \includegraphics[width=\textwidth]{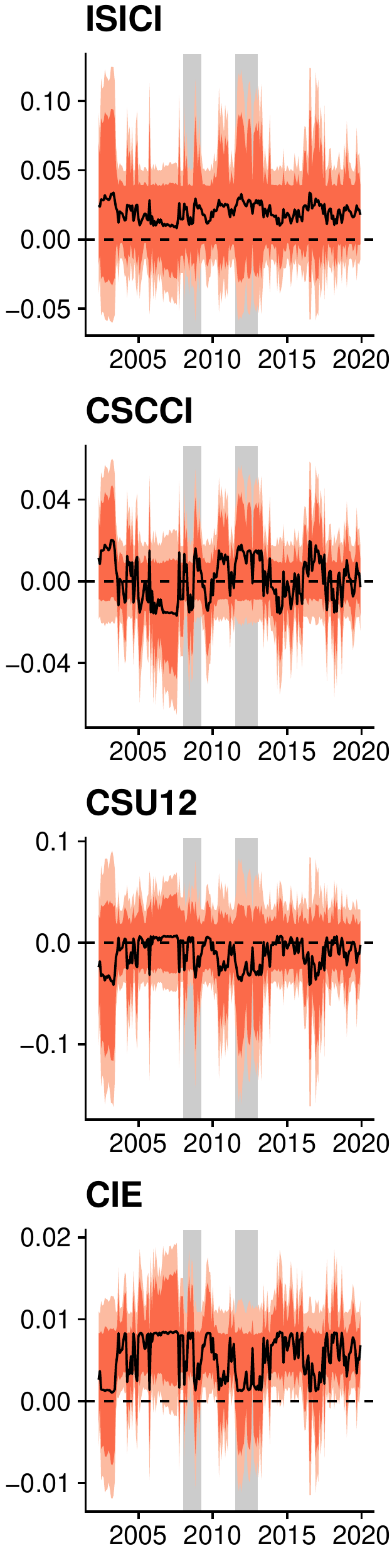}
        \caption{$h=4$}
    \end{subfigure}
    \begin{subfigure}[t]{0.19\textwidth}
        \includegraphics[width=\textwidth]{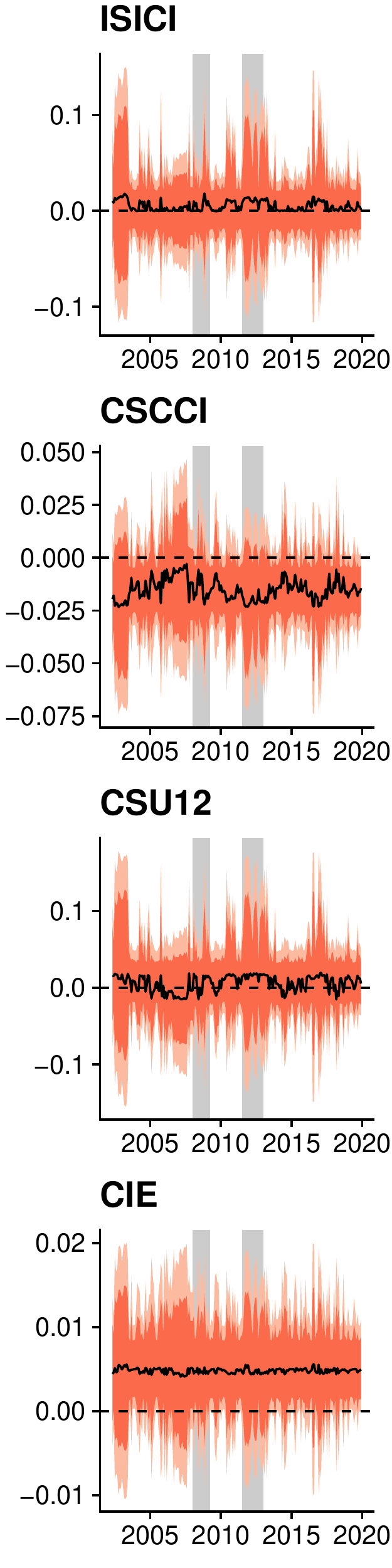}
        \caption{$h=12$}
    \end{subfigure}
    \begin{subfigure}[t]{0.19\textwidth}
        \includegraphics[width=\textwidth]{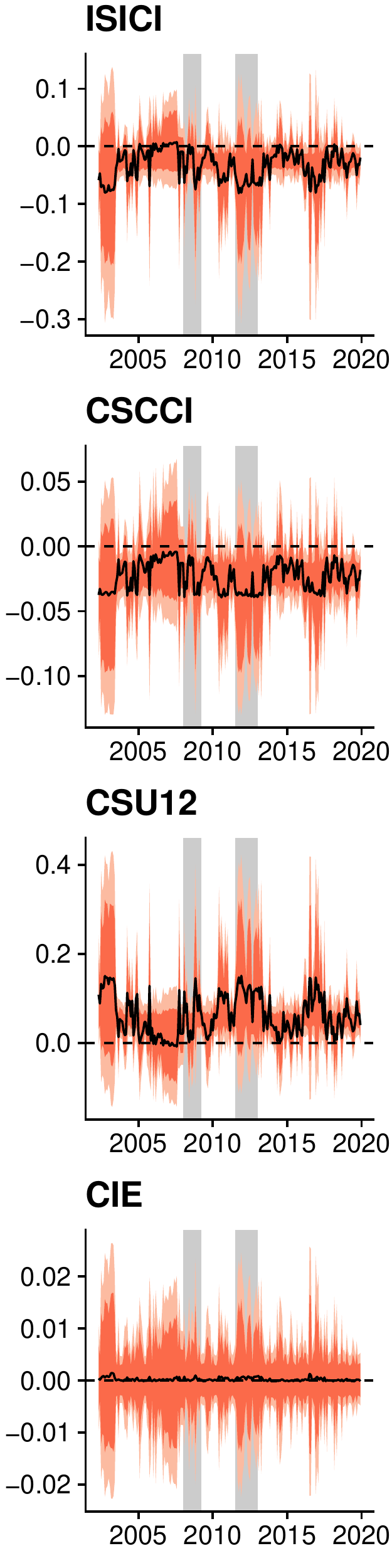}
        \caption{$h=24$}
    \end{subfigure}
    \begin{subfigure}[t]{0.19\textwidth}
        \includegraphics[width=\textwidth]{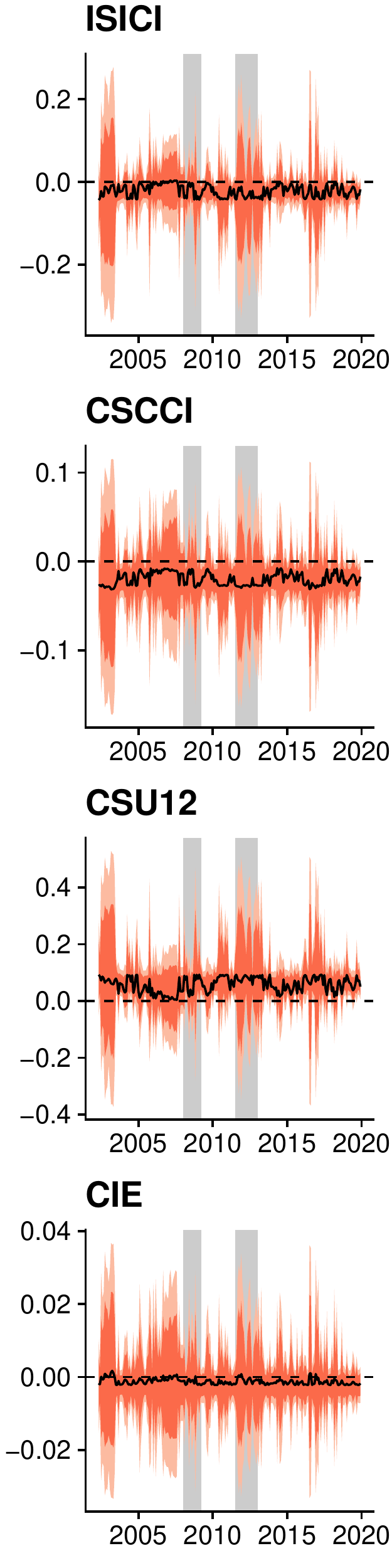}
        \caption{$h=36$}
    \end{subfigure}
    \caption{Responses of survey-based expectations to the FG shock at different horizons.}\label{fig:irfs_FG_survey}\vspace*{-0.3cm}
    \caption*{\footnotesize\textit{Notes}: The sample runs from 2002:01 to 2019:12. The horizontal dashed black line marks zero. The solid black line is the posterior median of the impulse response function for period $t$ at different horizons $h=\{0,4,12,24,36\}$, that is, the impact response, the one-quarter, one-year, two-years and three-years ahead response. The dark shades indicate the $50$ percent posterior credible set, the lighter shade the $68$ percent posterior credible set. The grey shaded areas indicate recessions dated by the CEPR Euro Area Business Cycle Dating Committee. For variable codes see Section \ref{sec:data} and Appendix \ref{app:C}.}
\end{figure}

\subsubsection{The quantitative easing shock}
Our final set of results is concerned with a contractionary QE shock. The impulse response functions are again scaled to reflect a one-standard deviation increase of the instrument. Since the respective QE factor is only active starting in 2014, we restrict our sample to the period where this policy measure is relevant. The results are shown in Figs. \ref{fig:irfs_QE_fin}, \ref{fig:irfs_QE_macro} and \ref{fig:irfs_QE_survey}.

Starting with financial quantities in Fig. \ref{fig:irfs_QE_fin}, we observe that by construction of the instrument, impact reactions of the short-term rate are muted and increase with longer maturity. The largest effects are around two basis points for the ten-year yields during periods of low uncertainty. Around the Brexit referendum in mid-2016, the shock does not translate into increases in interest rates, in line with our previous reasoning of impaired transmission during periods of elevated uncertainty. Similar to the FG shock and in contrast to TG, we observe insignificant reactions with respect to OAS, our variable measuring financial conditions. In line with the literature \citep[see, e.g.,][]{swanson2020measuring}, effects on the stock market are also muted.

\setcounter{figure}{0}
\renewcommand{\thefigure}{5.\arabic{figure}}
\begin{figure}[ht]
    \captionsetup[subfigure]{justification=centering}
    \begin{subfigure}[t]{0.19\textwidth}
        \includegraphics[width=\textwidth]{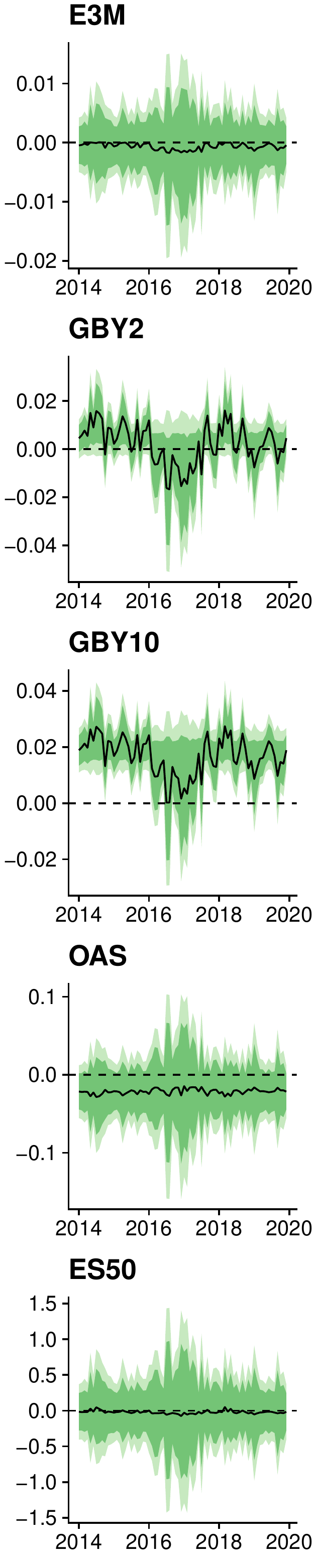}
        \caption{$h=0$}
    \end{subfigure}
    \begin{subfigure}[t]{0.19\textwidth}
        \includegraphics[width=\textwidth]{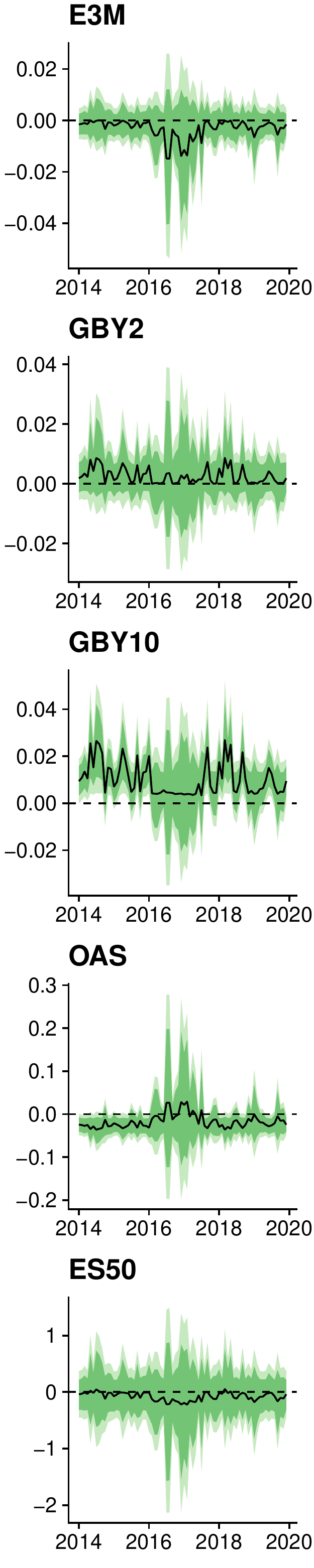}
        \caption{$h=4$}
    \end{subfigure}
    \begin{subfigure}[t]{0.19\textwidth}
        \includegraphics[width=\textwidth]{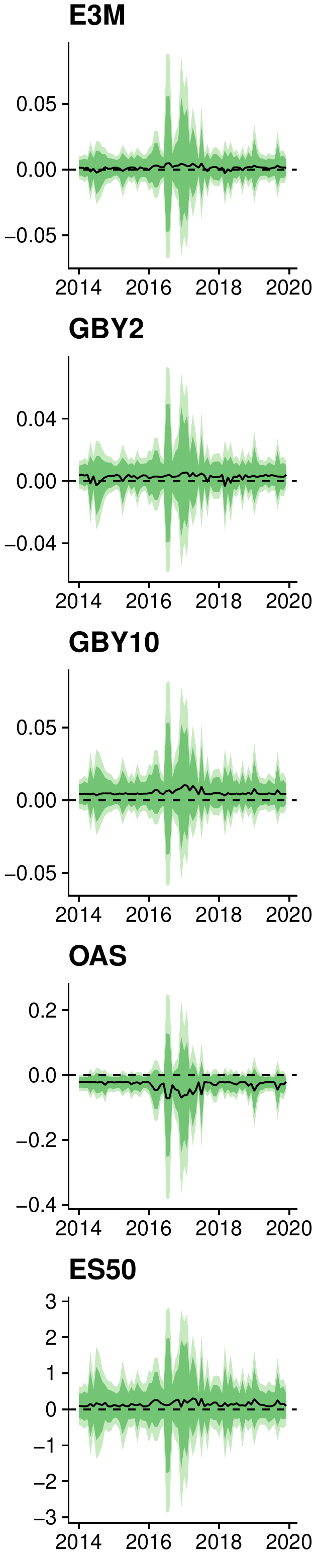}
        \caption{$h=12$}
    \end{subfigure}
    \begin{subfigure}[t]{0.19\textwidth}
        \includegraphics[width=\textwidth]{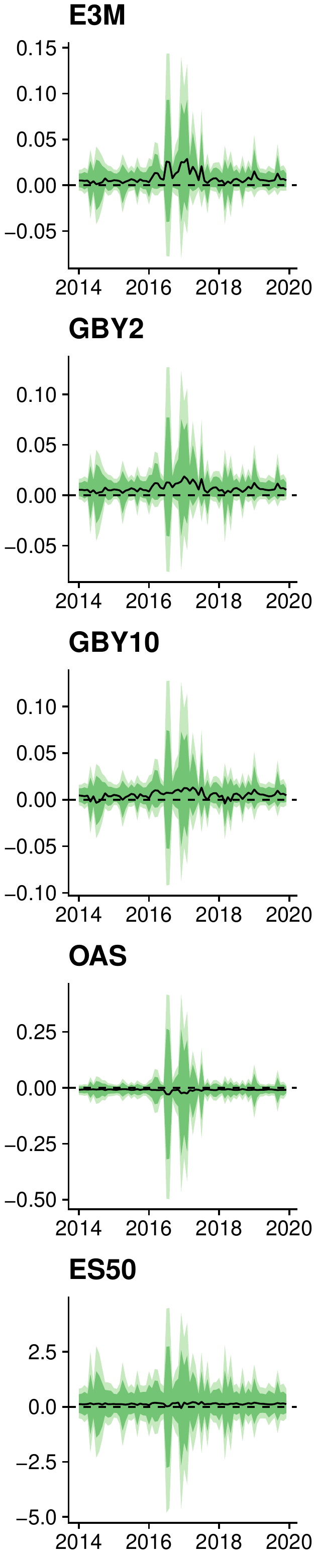}
        \caption{$h=24$}
    \end{subfigure}
    \begin{subfigure}[t]{0.19\textwidth}
        \includegraphics[width=\textwidth]{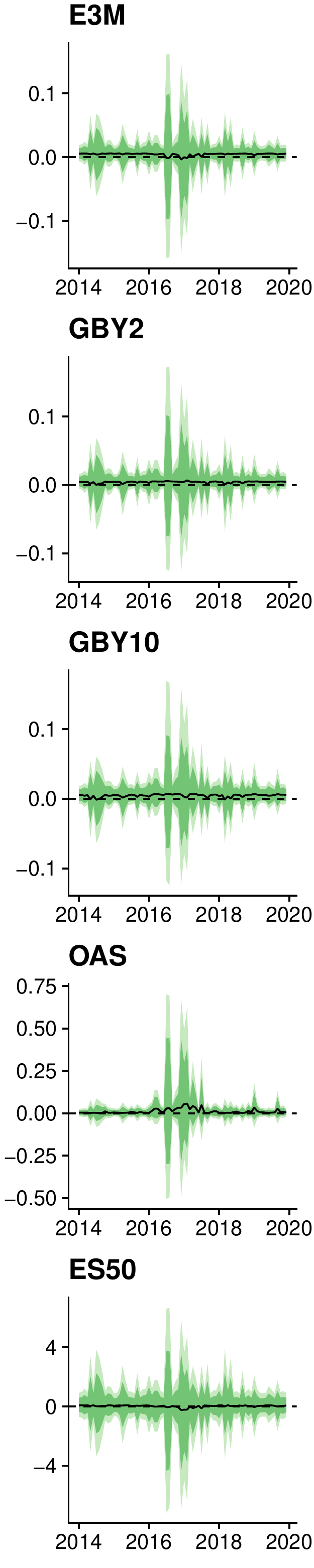}
        \caption{$h=36$}
    \end{subfigure}
    \caption{Responses of financial variables to the QE shock at different horizons.}\label{fig:irfs_QE_fin}\vspace*{-0.3cm}
    \caption*{\footnotesize\textit{Notes}: The sample runs from 2014:01 to 2019:12. The horizontal dashed black line marks zero. The solid black line is the posterior median of the impulse response function for period $t$ at different horizons $h=\{0,4,12,24,36\}$, that is, the impact response, the one-quarter, one-year, two-years and three-years ahead response. The dark shades indicate the $50$ percent posterior credible set, the lighter shade the $68$ percent posterior credible set. The grey shaded areas indicate recessions dated by the CEPR Euro Area Business Cycle Dating Committee. For variable codes see Section \ref{sec:data} and Appendix \ref{app:C}.}
\end{figure}

In terms of higher-order responses, we observe that all variables that do not show significant impact reactions do not turn significant at any horizon or any point in time. It is worth mentioning that the shocks are more short-lived when compared to FG, with two-year yield increases turning insignificant after one quarter, and ten-year yields returning to the baseline after about one year.

\begin{figure}[ht]
    \captionsetup[subfigure]{justification=centering}
    \begin{subfigure}[t]{0.19\textwidth}
        \includegraphics[width=\textwidth]{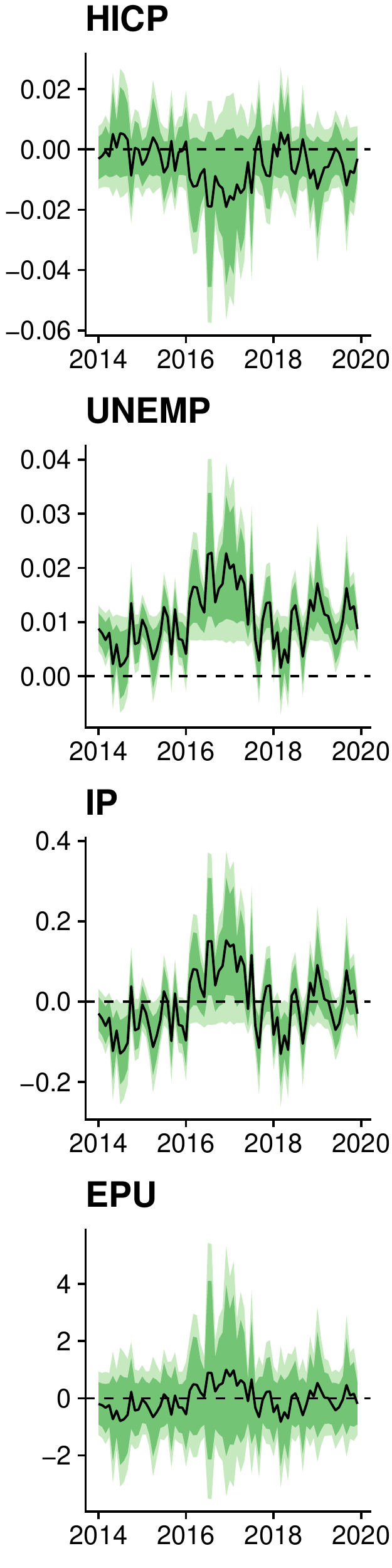}
        \caption{$h=0$}
    \end{subfigure}
    \begin{subfigure}[t]{0.19\textwidth}
        \includegraphics[width=\textwidth]{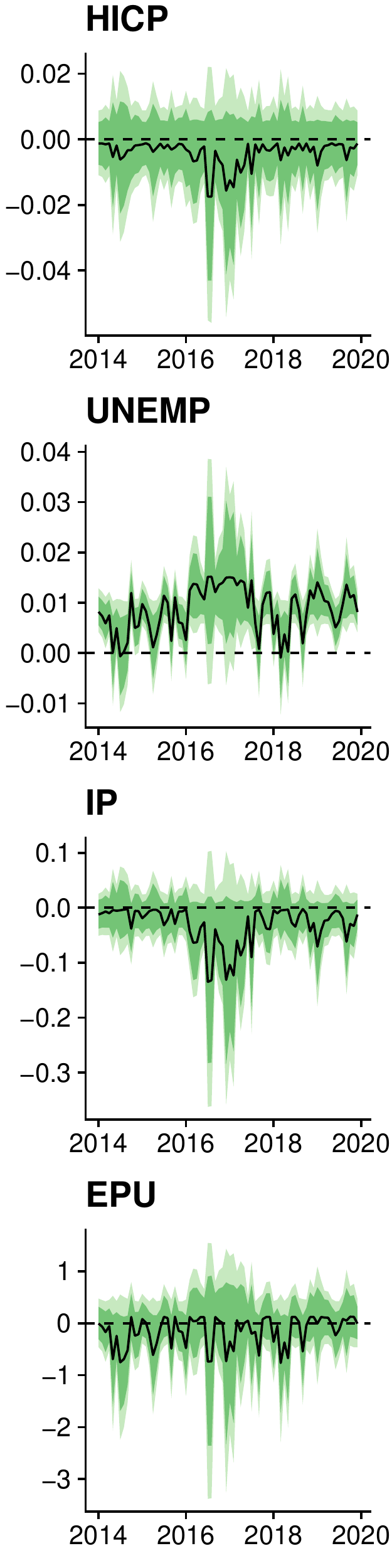}
        \caption{$h=4$}
    \end{subfigure}
    \begin{subfigure}[t]{0.19\textwidth}
        \includegraphics[width=\textwidth]{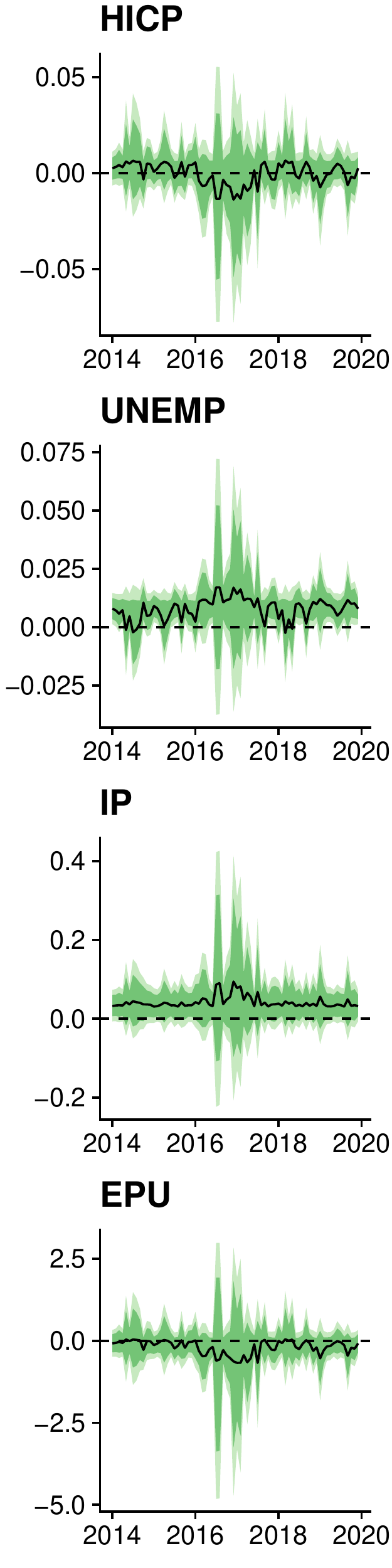}
        \caption{$h=12$}
    \end{subfigure}
    \begin{subfigure}[t]{0.19\textwidth}
        \includegraphics[width=\textwidth]{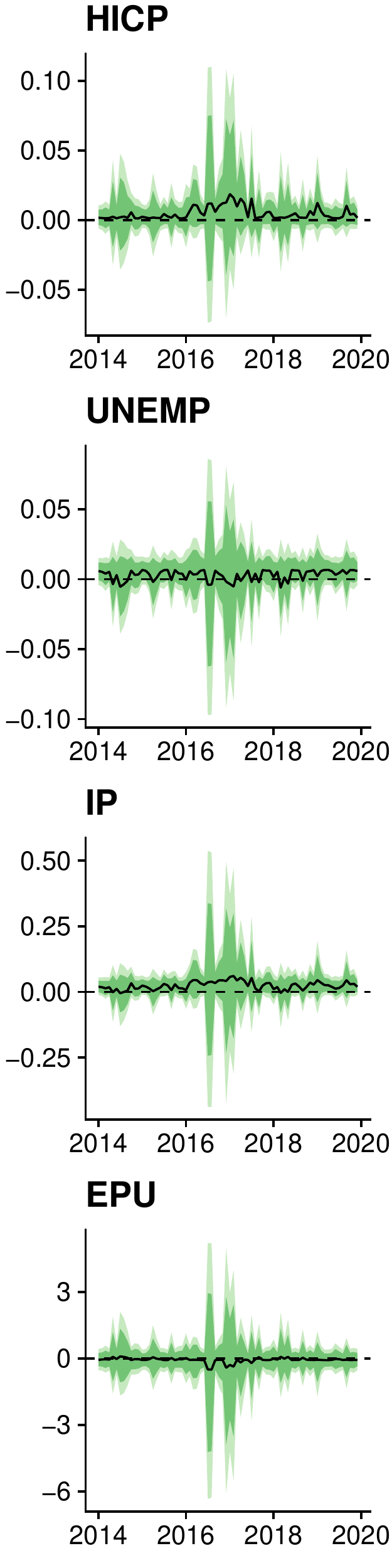}
        \caption{$h=24$}
    \end{subfigure}
    \begin{subfigure}[t]{0.19\textwidth}
        \includegraphics[width=\textwidth]{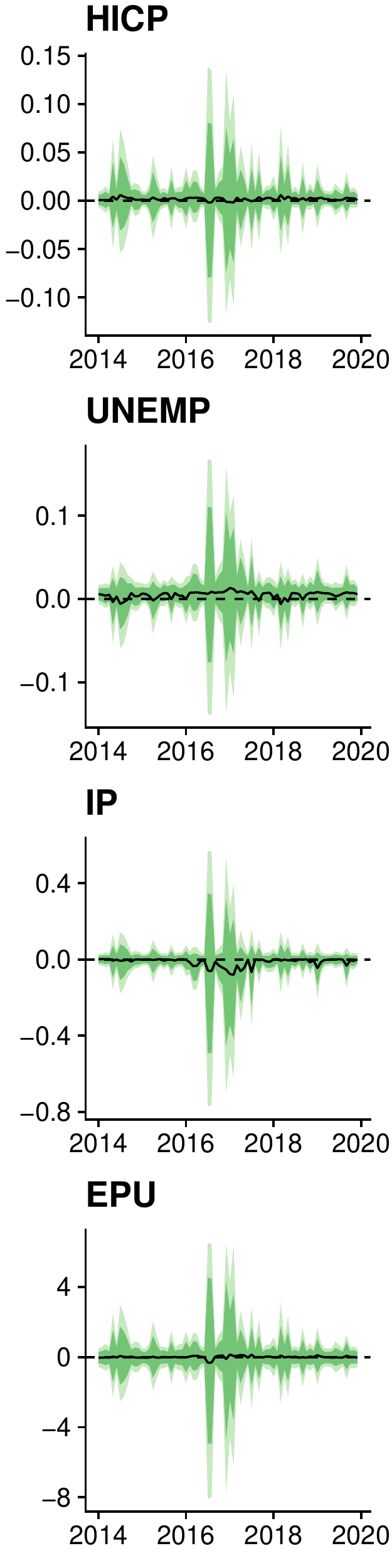}
        \caption{$h=36$}
    \end{subfigure}
    \caption{Responses of macroeconomic variables to the QE shock at different horizons.}\label{fig:irfs_QE_macro}\vspace*{-0.3cm}
    \caption*{\footnotesize\textit{Notes}: The sample runs from 2014:01 to 2019:12. The horizontal dashed black line marks zero. The solid black line is the posterior median of the impulse response function for period $t$ at different horizons $h=\{0,4,12,24,36\}$, that is, the impact response, the one-quarter, one-year, two-years and three-years ahead response. The dark shades indicate the $50$ percent posterior credible set, the lighter shade the $68$ percent posterior credible set. The grey shaded areas indicate recessions dated by the CEPR Euro Area Business Cycle Dating Committee. For variable codes see Section \ref{sec:data} and Appendix \ref{app:C}.}
\end{figure}

The results for our set of macroeconomic variables are displayed in Figure \ref{fig:irfs_QE_macro}. We detect minor differences in the responses conditional on the level of uncertainty for all series we consider. Interestingly, we find that estimated effects are a bit larger in magnitude if uncertainty is high. A contractionary shock translates to slight decreases of inflation (albeit insignificant) and substantial increases of unemployment. Economic activity (measured by industrial production) tends to decrease if uncertainty is low, with barely significant positive effects during mid to late 2016. The uncertainty index does not exhibit significant responses across impulse horizons and over time.

The minor puzzling result of positive industrial production responses when uncertainty is high reverses for one-quarter ahead impulses, albeit these estimates are not significant. Nevertheless, substantial posterior mass is allocated to negative values, particularly for periods of elevated uncertainty. Unemployment responses are comparatively persistent, with significant effects up to the one-year ahead horizon. 

Our discussion of the empirical results is completed by investigating the responses of the survey-based expectation measures in Fig. \ref{fig:irfs_QE_survey}. Similar to the macroeconomic variables, we find that for variables where we detect significant effects, they are largest in magnitude during periods of elevated uncertainty. A contractionary QE shock results in decreases particularly around 2016 for industry confidence. Impact reactions for consumer confidence are insignificant, but turn significant for the one-quarter ahead horizon. The timing of the response is also interesting for unemployment expectations. While impact reactions are stronger during low-uncertainty periods and insignificant for high uncertainty, this changes at the one-quarter ahead horizon with substantially larger effects in times of uncertainty. Similar as in the context of the FG shock, we detect a puzzle for consumer inflation expectations, with positive effects on impact that peter out quickly after one quarter. It is worth mentioning that the QE shock results in less persistent impulse responses, with all effects turning insignificant at the latest after one year.

Summarising our results on UMP, we find that both FG and QE shocks exhibit expected effects on interest rates of different maturity. While the effects of the FG shock seem to be independent of the prevailing level of uncertainty, effects of QE are more strongly translated into interest rates in episodes of higher uncertainty. Similarly, QE effects on the real economy and expectations seem to be greater when uncertainty is elevated. Here, both the portfolio rebalancing channel as well as the policy signalling channel may play a role in the transmission of QE \citep[see][]{hutchinsonsmets2017}. Our results imply that this instrument appears to be particularly suited for periods characterized by high levels of uncertainty. 

\begin{figure}[ht]
    \captionsetup[subfigure]{justification=centering}
    \begin{subfigure}[t]{0.19\textwidth}
        \includegraphics[width=\textwidth]{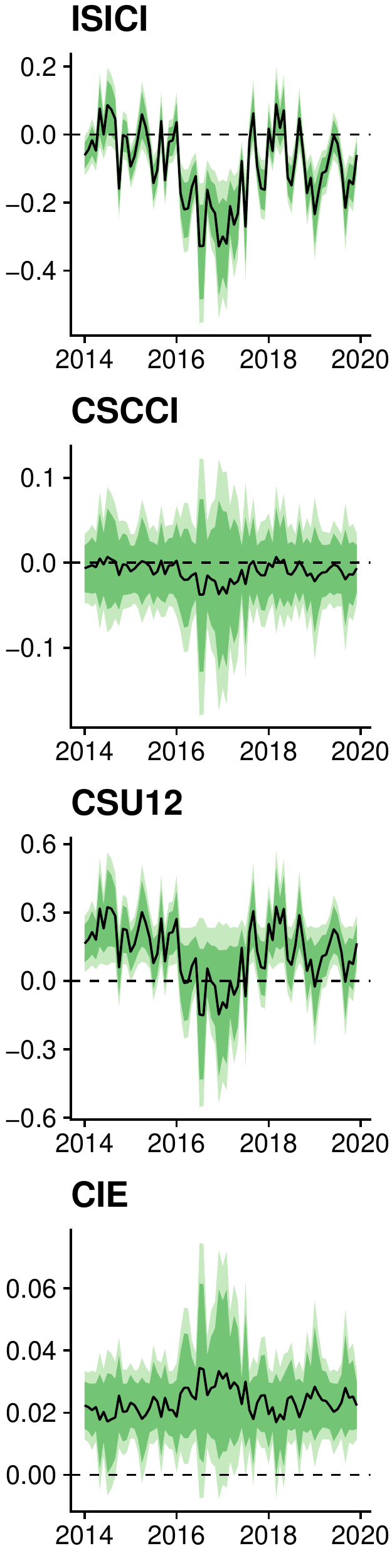}
        \caption{$h=0$}
    \end{subfigure}
    \begin{subfigure}[t]{0.19\textwidth}
        \includegraphics[width=\textwidth]{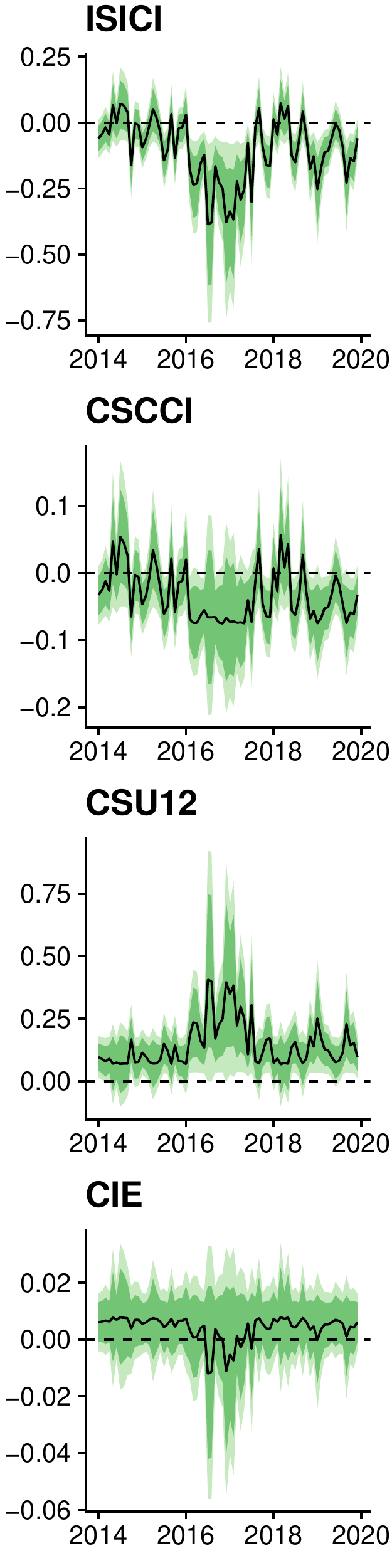}
        \caption{$h=4$}
    \end{subfigure}
    \begin{subfigure}[t]{0.19\textwidth}
        \includegraphics[width=\textwidth]{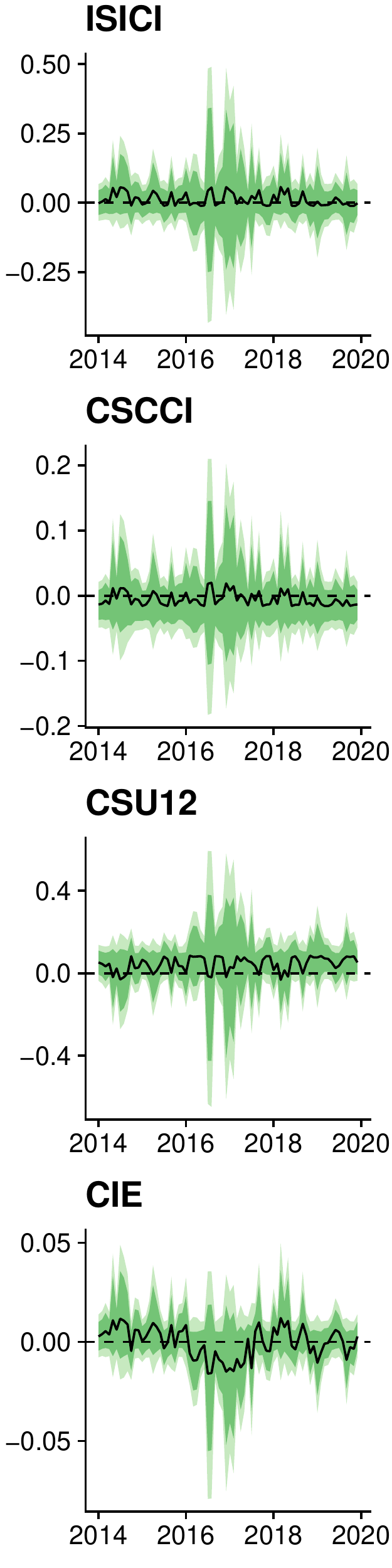}
        \caption{$h=12$}
    \end{subfigure}
    \begin{subfigure}[t]{0.19\textwidth}
        \includegraphics[width=\textwidth]{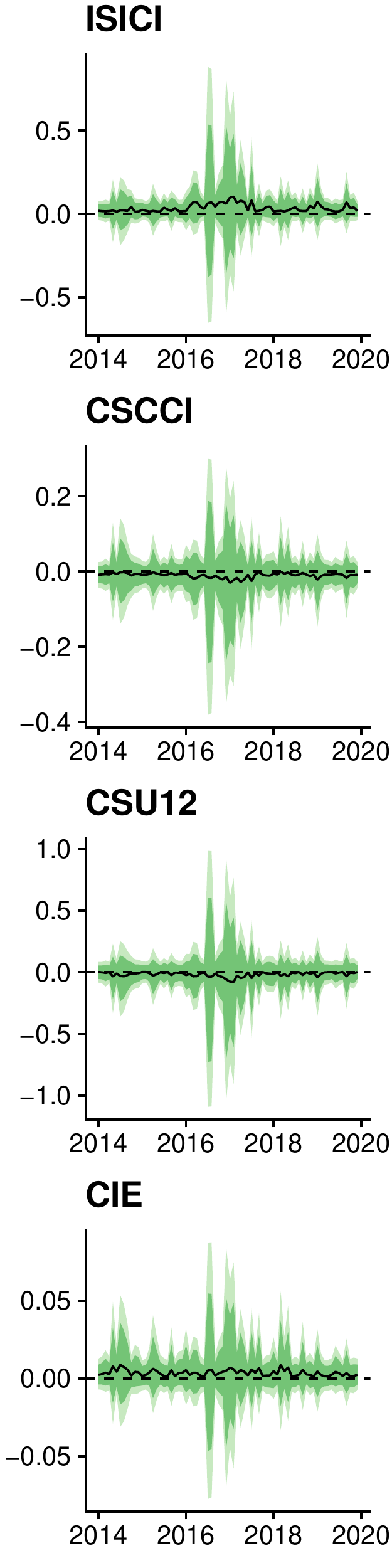}
        \caption{$h=24$}
    \end{subfigure}
    \begin{subfigure}[t]{0.19\textwidth}
        \includegraphics[width=\textwidth]{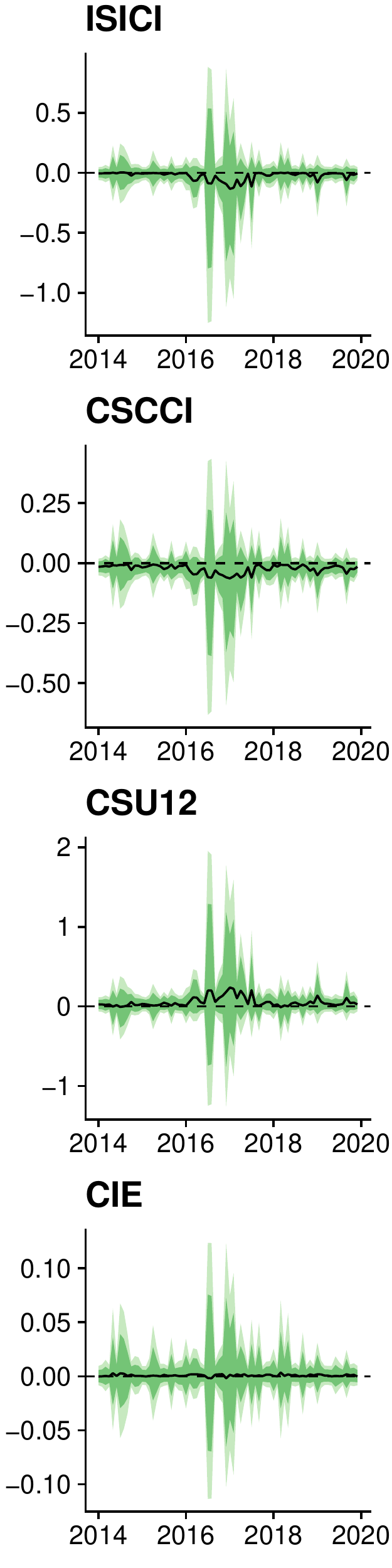}
        \caption{$h=36$}
    \end{subfigure}
    \caption{Responses of survey-based expectations to the QE shock at different horizons.}\label{fig:irfs_QE_survey}\vspace*{-0.3cm}
    \caption*{\footnotesize\textit{Notes}: The sample runs from 2014:01 to 2019:12. The horizontal dashed black line marks zero. The solid black line is the posterior median of the impulse response function for period $t$ at different horizons $h=\{0,4,12,24,36\}$, that is, the impact response, the one-quarter, one-year, two-years and three-years ahead response. The dark shades indicate the $50$ percent posterior credible set, the lighter shade the $68$ percent posterior credible set. The grey shaded areas indicate recessions dated by the CEPR Euro Area Business Cycle Dating Committee. For variable codes see Section \ref{sec:data} and Appendix \ref{app:C}.}
\end{figure}
\setcounter{figure}{5}
\renewcommand{\thefigure}{\arabic{figure}}

\section{Closing remarks}\label{sec:conclusions}
In this paper, we assess the effectiveness of several conventional and unconventional monetary policy measures by the European Central Bank conditional on the prevailing level of uncertainty. We measure effects of target, forward guidance and quantitative easing shocks over time with a smooth-transition vector autoregressive model using uncertainty as a signal variable. This allows for obtaining time-varying impulse response functions of a set of financial, macroeconomic and survey-based expectation variables to the shocks.

Our results suggest that transmission channels are impaired in times of uncertainty. We conjecture that this is due to two major reasons. First, high levels of uncertainty change the impact effects of the shocks by changing direct transmission to key financial variables such as spreads. We argue that this phenomenon may be related to the presence of central bank information effects during these periods that accompany monetary policy announcements. Second, we find that the prevailing level of uncertainty at the time of the shock affects the persistence of the effects. The persistence of the shocks is dependent on the specific instrument, but is also variable specific. Our results suggest that especially the effectiveness of transmission channels related to industrial and consumer expectations are affected in times of uncertainty.

The effectiveness of policy measures also depends crucially on the respective tool invoked by the central bank. Conventional policy such as shocks to the target rate appear to be particularly affected negatively by high uncertainty. By contrast, quantitative easing measures such as large-scale asset purchases appear to work particularly well in times of high uncertainty. For both UMP measures, our analysis suggests that there might be beneficial complementarities between TG, FG and QE. On the one hand, FG might be useful in combination with a CMP instrument, making its effects more persistent and more robust with respect to different uncertainty levels. On the other hand, a QE shock in combination with CMP might not only add to the effects of a change in the target rate via the portfolio rebalancing and signalling channel, but it might also make up for deficiencies in the transmission of CMP in times of elevated uncertainty.

\clearpage\small{\setstretch{0.85}
\addcontentsline{toc}{section}{References}
\bibliographystyle{custom.bst}
\bibliography{lit}}\normalsize

\clearpage
\begin{appendices}\crefalias{section}{appsec}
\begin{center}
{\Large\textbf{Appendix}}
\end{center}

\setcounter{equation}{0}
\renewcommand\theequation{A.\arabic{equation}}
\section{Equation-by-equation estimation}\label{app:A}
In this Appendix we outline the estimation of our smooth transition VAR model in \autoref{eq:benchmarkVAR}. For notational simplicity, we first collect the respective coefficient matrices in an $M \times K$-matrix, where $K = 2(MP+2)$, and the regressors (i.e., lagged dependent variables and the exogenous instrument) in a $K \times 1$-vector:
\begin{align*}
    \bm A =& (\bm{A}_{11}, \hdots, \bm{A}_{1P}, \bm{c}_1, \bm \delta_1, \bm{A}_{01}, \hdots, \bm{A}_{0P}, \bm{c}_0, \bm \delta_0), \\
    \bm z_t =& \left((\bm{y}'_{t-1},  \hdots \bm{y}'_{t-P}, 1, x_{st}) \times S_t(u_{t-1}), (\bm{y}'_{t-1}, \hdots \bm{y}'_{t-P}, 1, x_{st}) \times (1-S_t(u_{t-1}))\right)'.
\end{align*}
We rewrite \autoref{eq:benchmarkVAR} more compactly as follows: 
\begin{align*}
    \bm{y}_t = \bm A \bm z_t  + \bm H \bm \eta_t, \quad \bm{\eta}_t \sim \mathcal{N}(\bm 0, \bm \Sigma),
\end{align*}
with $\bm \eta_t$ denoting structural errors. Note that the relation between reduced-form errors $\bm \epsilon_t$ and structural errors $\bm \eta_t $ is given by $\bm \epsilon_t = \bm H \bm \eta_t$. 

By exploiting the fact that $\bm \eta_t = \bm H^{-1} \bm \epsilon_t$, and the lower triangular structure of $\bm H^{-1}$, we estimate the system equation-by-equation as a set of unrelated regressions, see \cite{carriero2019large}. The first equation in \autoref{eq:benchmarkVAR} is:
\begin{align*}
     y_{1t} = \bm A_{1\bullet} \bm z_t + \eta_{1t}, \quad \eta_{1t} \sim \mathcal{N}(0, \sigma^2_{1}),  
\end{align*}
where $\bm A_{1\bullet}$ refers to the first row in $\bm A$. The $m$th (for $m = 2, \hdots, M$) equation of \autoref{eq:benchmarkVAR} is: 
\begin{align*}
     y_{mt} = \bm A_{m \bullet} \bm z_t  - \sum_{i = 1}^{m-1} h_{mi}^{-1} \epsilon_{it} + \eta_{mt}, \quad \eta_{mt} \sim \mathcal{N}(0, \sigma^2_{m}),
\end{align*}
where $\bm A_{m \bullet}$ refers to the $m$th row in $\bm A$ and $h_{mi}^{-1}$ to the $(m,i)$th element in $\bm H^{-1}$. To simplify the $m$th equation, we define the $K_m \times 1$-vectors $\bm \alpha_{m} = (\bm A_{m \bullet}', \{h_{mi}^{-1}\}_{i =1}^{m-1})'$ and $\tilde{\bm z}_{mt} = (\bm z_t', \{\epsilon_{it}\}_{i =1}^{m-1})'$ and write 
\begin{align}
     y_{mt} = \bm \alpha_{m}' \tilde{\bm z}_{mt} + \eta_{mt}, \quad \eta_{mt} \sim \mathcal{N}(0, \sigma^2_{m}).\label{eq:eqbyeq}
\end{align}
Note that, if $m = 1$, $\bm \alpha_{1} = A_{1\bullet}$ and $\tilde{\bm z}_{1t} = \bm z_t$. To define the \autoref{eq:eqbyeq} in terms of full data matrices, let $\bm y_m$ be a $T \times 1$-vector, $\tilde{\bm Z}_m$ a $T \times K_i$ matrix and $\bm \eta_{m}$ a $T \times 1$-vector, collecting $y_{mt}$, $\tilde{\bm z}'_{mt}$ and $\eta_{mt}$ on the $t$th position, respectively. The following equation, conditional on $S_t(u_{t-1})$, denotes a standard linear regression model: 
\begin{align*}
     \bm y_{m} = \tilde{\bm Z}_{m}\bm \alpha_{m} + \bm \eta_{m}, \quad \bm \eta_{m} \sim \mathcal{N}(\bm 0, \sigma^2_{m} \bm I_T).
\end{align*}

\clearpage\renewcommand\theequation{B.\arabic{equation}}
\section{Posterior distributions and sampling algorithm}\label{app:B}
In this Appendix we briefly summarize the main steps involved for posterior inference. Given the model likelihood and prior assumptions we can derive a standard Markov Chain Monte Carlo (MCMC) algorithm, which iterates through conditional posterior distributions. The symbol $\star$ indicates that we condition on all other parameters of the model (including the state indicator $S_t(u_{t-1})$).

\begin{enumerate}[leftmargin=*]
    \item We draw $\bm \alpha_{m}$, for $m = 1, \dots, M$, from a multivariate Normal distribution:
    \begin{equation*}
        \bm \alpha_m|\star = \mathcal{N}(\bar{\bm \alpha}_m, \bar{\bm \Omega}_m),
    \end{equation*}
    with $\bar{\bm \alpha}_m$ denoting the posterior mean and $\bar{\bm \Omega}_m$ refering to the posterior variance-covariance matrix. Both quantities are of well-known form an given by: 
    \begin{align*}
     \bar{\bm \alpha}_m =& \bar{\bm V}_m(\tilde{\bm Z}_{m}'\bm y_{m}/\sigma^2_{m} + \underline{\bm \Omega}_m^{-1} \underline{\bm \alpha}_{m}), \\
     \bar{\bm \Omega}_m =& (\tilde{\bm Z}_{m}'\tilde{\bm Z}_{m}/\sigma^2_{m} + \underline{\bm \Omega}_m^{-1})^{-1},
    \end{align*}
    with the $K_m \times 1$-vector $\underline{\bm \alpha}_{m}$ being a prior mean and the $K_m \times K_m$-matrix  $\underline{\bm \Omega}_m^{-1}$ a diagonal prior variance covariance matrix. In the following, the prior quantities collect the equation-specific elements defined by the hierarchical prior. That is, $\underline{\bm \alpha}_{m}$ collects the respective elements of $\tilde{\bm a}$, defined in Eq. (\ref{eq:poolVAR}), elements of the lower Cholesky factor are centered on zero from Eq. (\ref{eq:covVAR}), $\underline{\bm \Omega}_m^{-1}$ collects the respective elements of $\{\tilde{v}_{1j}\}_{j=1}^{J}$, $\{\tilde{v}_{0j}\}_{j=1}^{J}$ and $\{\hat{v}_{0j}\}_{j=1}^{J}$.\footnote{In particular, for the $m$th equation, $\underline{\bm \alpha}_{m} = (\{\tilde{a}_j\}_{j \in S_m}, \{\tilde{a}_j\}_{j \in S_m}, \bm 0_{m-1}')'$, where $\tilde{a}_{j}$ denotes the $j$th element in $\tilde{\bm a}$ and $\bm 0_{m-1}$ refers to $(m-1)\times 1$-vector of zeros, while $\underline{\bm \Omega}_m^{-1} = \text{diag}(\{\tilde{v}_{1s}\}_{s \in \tilde{S}_m}, \{\tilde{v}_{0s}\}_{s \in \tilde{S}_m}, \{\hat{v}_{q}\}_{q \in Q_m})$. Here, $\tilde{S}_m = \{k(m-1)+1,\hdots,km\}$ is a set of indicators of cardinality $k = (MP+2)$ that serves to select the equation-specific coefficients and the set $Q_m$ of cardinality $(m-1)$ selects the prior indicators corresponding to equation-specific elements of the lower Cholesky factor.}
    \item We sample the structural error variances $\{\sigma^2_m\}_{m = 1}^{M}$ from a inverse Gamma distribution: 
    \begin{equation*}
        \sigma_m|\star \sim \mathcal{G}^{-1}(d_m, D_m),
    \end{equation*}
    with $d_m = (T/2 + 3)$ denoting the posterior degrees of freedom and $D_m = (\bm \eta_m'\bm \eta_m/2 + 0.3)$ refering to the posterior scaling parameter. 
    \item To update the hierarchical priors we rely on steps outlined next. First, note that that Eq. (\ref{eq:regimeVAR}) and Eq. (\ref{eq:poolVAR}) can be written as a random coefficient specification for each parameter. That is, 
    \begin{align*}
        a_{ij} &= \tilde{a}_j + \tilde{\nu}_{ij}, \quad \tilde{\nu}_{ij} \sim \mathcal{N}(0, \tilde{v}_{ij}) \\  
        \tilde{a}_j &= a_j + \nu_j, \quad \nu_j \sim \mathcal{N}(0, v_{j}).
    \end{align*}
    with $i \in \{0, 1\}$ and $j = 1, \dots, J$. 
    
    The (hyper)parameters of the hierarchical shrinkage prior are sampled from the following conditional posterior distributions: 
    \begin{itemize}[leftmargin = 1.5em]
        \item The HS scaling parameters of the lowest hierarchy (i.e., $\tilde{v}_{j} = \tilde{\lambda}_i^2\tilde{\psi}_{ij}^2$) are obtained by using the methods outlined in \cite{makalic2015simple}. That is, for $i \in \{0, 1\}$ and $j = 1, \dots, J$, sampling the local $\tilde{\psi}_{ij}^2$ and the global shrinkage parameter $\tilde{\lambda}_i^2$ involves producing draws from four independent inverse Gamma distributions. 
        For the $\tilde{\psi}_{ij}^2$ the distribution is given by: 
        \begin{equation*}
        \tilde{\psi}^2_{ij}|\star \sim \mathcal{G}^{-1}\left(1, \frac{1}{\tilde{\zeta}_{ij}}+\frac{(a_{ij} - \tilde{a}_j)^2}{2\tilde{\lambda}_i^2}\right),
        \end{equation*}
        the global shrinkage parameter is sample from: 
        \begin{equation*}
        \tilde{\lambda}_i^2|\star \sim \mathcal{G}^{-1}\left(\frac{J+1}{2}, \frac{1}{\tilde{\xi}_i}+ \sum_{j=1}^{J}\frac{(a_{ij} - \tilde{a}_j)^2}{2\tilde{\psi}_{ij}^2}\right),
        \end{equation*}
        while the two auxiliary variables $\tilde{\zeta}_{ij}$ and $\tilde{\xi}_{i}$ are drawn from: 
        \begin{equation*}
        \begin{aligned}
            \tilde{\zeta}_{ij}|\star & \sim  \mathcal{G}^{-1}(1, 1 + 1/\tilde{\psi}^2_{ij}), \\
            \tilde{\xi}_{i}|\star & \sim  \mathcal{G}^{-1}(1, 1 + 1/\tilde{\lambda}_i^2).
        \end{aligned}   
        \end{equation*}
    \item Similar to the lowest hierarchy we also defined a HS prior on the top one. To sample the scaling parameters $\psi^2_{j}$ (local scaling) and $\lambda^2$ (global scaling), we therefore rely on exactly the same steps outlined in the four equations above. This can be done by replacing  $(a_{ij} - \tilde{a}_j)^2$ with $(\tilde{a}_j - a_j)^2$, $\tilde{\lambda}_i$ with $\lambda$ and $\tilde{\zeta}_{ij}$ with $\zeta_{j}$ in the respective equations.
    
    \item In a next step, we update the hierarchical (common) prior mean $\tilde{\bm{a}}$ element-wise from a Gaussian distribution:
    \begin{equation*}
    \tilde{a}_j|\star \sim \mathcal{N}(\overline{a}_j, \overline{v}_j),
    \end{equation*}
    for $j = 1, \dots, J$. 
    The posterior mean $\overline{a}_j$ and variance $\overline{v}_j$ are given by: 
    \begin{align*}
    \overline{v}_j &= \left(\left(\frac{1}{\tilde{v}_{1j}} + \frac{1}{\tilde{v}_{0j}}\right) + \frac{1}{v_j} \right)^{-1}, \\
    \overline{a}_j &=  \overline{v}_j \left(\left(\frac{a_{1j}}{\tilde{v}_{1j}} + \frac{a_{0j}}{\tilde{v}_{0j}}\right) + \frac{a_j}{v_j} \right).
    \end{align*}    
    \end{itemize}
    \item The HS hyperparameters of elements associated with the covariances are sampled from:  
        \begin{equation*}
         \hat{\psi}^2_{r}|\star \sim \mathcal{G}^{-1}\left(1, \frac{1}{\hat{\zeta}_{r}}+\frac{h_r^2}{2\hat{\lambda}^2}\right),
        \end{equation*}
        the global shrinkage parameter is sampled from: 
        \begin{equation*}
        \hat{\lambda}^2|\star \sim \mathcal{G}^{-1}\left(\frac{R+1}{2}, \frac{1}{\hat{\xi}}+ \sum_{r=1}^{R}\frac{h_r^2}{2 \hat{\psi}_{r}^2}\right),
        \end{equation*}
        while the two auxiliary variables are drawn from: 
        \begin{equation*}
        \begin{aligned}
            \hat{\zeta}_{r}|\star & \sim \mathcal{G}^{-1}(1, 1 + 1/\hat{\psi}^2_{r}), \\
            \hat{\xi}|\star & \sim  \mathcal{G}^{-1}(1, 1 + 1/\hat{\lambda}^2).
        \end{aligned}   
        \end{equation*}
    \item Finally, following \citet{alessandrimumtaz2019} (see their Appendix for the ST-VAR), we sample the threshold parameters in one block using a random walk Metropolis Hastings step. We therefore use two independent Gaussian proposal densities to sample candidate values for $\gamma$ and $\phi$: 
    \begin{equation*}
      \gamma^{(*)} \sim \mathcal{N}(\gamma^{(s)}, c_{\gamma}), \quad \phi^{(*)} \sim \mathcal{N}(\phi^{(s)}, c_{\phi}).
    \end{equation*}
    Here, $\gamma^{(*)}$ refers to the candidate value and $\gamma^{(s)}$ to the last accepted draw of $\gamma$, while $\phi^{(*)}$ denotes the proposed value and $\phi^{(s)}$ the last accepted draw of $\phi$. 
    In the following, the acceptance probability $\omega$ is given by the ratio of the posterior likelihood of the proposed values ($\gamma^{(*)}, \phi^{(*)}$) and the posterior likelihood of the last accepted values ($\gamma^{(s)}, \phi^{(s)}$), since both proposal are symmetric. That is,
   \begin{equation*}
        \omega = \min \left(1, \frac{ \mathcal{L}(\gamma^{(*)}, \phi^{(*)}|\star) p(\gamma^{(*)}) p(\phi^{(*)})}{\mathcal{L}(\gamma^{(s)}, \phi^{(s)}|\star) p(\gamma^{(s)}) p(\phi^{(s)})} \right), 
    \end{equation*}
    with $\mathcal{L}$ denoting the conditional data likelihood. Note that the prior of $\gamma$ is defined as a truncated Gaussian distribution implying that the acceptance probability of the hyperparameter pair ($\gamma^{(*)}, \phi^{(*)}$) is zero, if $\gamma^{(*)} < \min(z_t)$ or $ \gamma^{(*)} > \max(z_t)$ as these values obtain zero support per construction.  In the empirical application, we choose $c_{\gamma}$ and $c_{\phi}$ in such a way to obtain an acceptance rate between $25$ to $40$ percent.
\end{enumerate}
We repeat these steps $32,000$ times and discard the initial $2,000$ draws as a burn-in. We consider each $10$th of the retained draws, resulting in a set of $3,000$ independent draws from the posterior for inference.

\clearpage
\renewcommand\theequation{C.\arabic{equation}}
\section{Data sources and transformations}\label{app:C}

\begin{table*}[h]
\caption{Dataset.}\vspace*{-1.5em}
\begin{center}
\begin{small}
\begin{threeparttable}
\begin{tabular*}{\textwidth}{@{\extracolsep{\fill}} llll}
  \toprule
  \textbf{Variable} & \textbf{Description} & \textbf{Source} & \textbf{Trans.}\\ 
  \midrule
  & \multicolumn{3}{l}{\textit{Financial variables}} \\
  \cmidrule{2-4}
  \texttt{E3M} & Euribor 3-month rate, monthly percentage & SDW & \\ 
  \texttt{GBY2} & EA 2-year government bond yield, monthly percentage & SDW & \\ 
  \texttt{GBY10} & EA 10-year government bond yield, monthly percentage & SDW & \\ 
  \texttt{ES50} & Euro Stoxx 50, price index, monthly & SDW & $100\cdot$log$(x)$\\
  \texttt{OAS}  & ICE BofA Euro high yield index & FRED\\
       &  option-adjusted spread (OAS), monthly percentage & & \\
  \midrule
  & \multicolumn{3}{l}{\textit{Macroeconomic variables}} \\
  \cmidrule{2-4}
  \texttt{HICP} & Harmonised index of consumer prices & SDW & $100\cdot$log$(x)$, yoy\\
       & monthly index, w.d.a, s.a. & & \\
  \texttt{UNEMP} & Harmonized Unemployment Rate & FRED&\\
       & monthly, s.a. & \\
  \texttt{IP} & Industrial production, excl. construction & SDW & $100\cdot$log$(x)$, yoy\\
       & monthly, w.d.a, s.a. &  & \\
  \texttt{EPU}  & Economic policy uncertainty, monthly & BBD & $100\cdot$log$(x)$ \\
  \midrule
  & \multicolumn{3}{l}{\textit{Expectations/survey-based variables}} \\
  \cmidrule{2-4}
  \texttt{ISICI} & Industrial confidence indicator, monthly percentage & SDW &\\ 
  \texttt{CSCCI} & Consumer confidence indicator, monthly percentage & SDW &\\ 
  \texttt{CSU12} & Consumer unemployment expectations & SDW &\\ 
        & over next 12 months, monthly, percentage  & & \\
  \texttt{CIE}  & Consumer opinion future tendency of inflation & FRED & $x/10^{\dagger}$\\
       & monthly, s.a. & & \\
\bottomrule
\end{tabular*}
\begin{tablenotes}[para,flushleft]
\scriptsize{\textit{Notes}: Column \textit{Trans.} indicates the transformation applied to the respective series $x$. $^{\dagger}$Transformed to correspond to the scale of HICP inflation. FRED indicates the database maintained by the Federal Reserve Bank of St. Louis (\href{https://fred.stlouisfed.org}{fred.stlouisfed.org}), SDW is the statistical data warehouse by the European Central Bank (ECB, \href{https://sdw.ecb.europa.eu}{sdw.ecb.europa.eu}). The abbreviation \textit{s.a.} is short for seasonally adjusted, \textit{w.d.a.} means working-day adjusted (both only stated if applicable), \textit{yoy} refers to year-on-year differenced data. \textit{BBD} is short for \citet{baker2016measuring}, who provide the economic policy uncertainty index on their webpage: \href{http://www.policyuncertainty.com}{policyuncertainty.com}.}
\end{tablenotes}
\end{threeparttable}
\end{small}
\end{center}
\label{tab:data}
\end{table*}

\end{appendices}
\end{document}